\documentclass[twoside,leqno,twocolumn]{article}

\usepackage[letterpaper]{geometry}

\usepackage{ltexpprt}
\usepackage{hyperref}
\usepackage{comment}
\usepackage{lipsum}
\usepackage{rotating} 
\usepackage{geometry}
\usepackage{amsmath}
\usepackage{amssymb}
\usepackage{wasysym}
\usepackage{algorithm}
\usepackage{algorithmic}
\usepackage{graphics}
\usepackage{graphicx}
\usepackage{float}
\usepackage{xspace}
\usepackage{xcolor}
\usepackage{caption}
\usepackage{fancyhdr}
\usepackage{booktabs}
\usepackage[section]{placeins}
\usepackage[T1]{fontenc}

\newcommand{\BaseInfer}{\textsc{BaseInfer}}
\newcommand{\MDLFramework}{\textsc{MdlInfer}}
\newcommand{\MDLResult}{\textsc{MdlParam}\xspace}
\newcommand{\MordecaiResult}{\textsc{BaseParam}\xspace}

\newcommand{\SAPHIRE}{\mathrm{SAPHIRE}}
\newcommand{\SEIR}{\mathrm{SEIR+HD}}
\newcommand{\MCMC}{\mathrm{MCMC}}
\newcommand{\IteratedFiltering}{\mathrm{IF}}
\newcommand{\Calibrate}{\textsc{Calibrate}}
\newcommand{\Cost}{\textsc{Cost}}
\newcommand\Vector[1]{\textbf{#1}}
\newcommand{\hide}[1]{}

\begin{document}

\newcommand\relatedversion{}
\renewcommand\relatedversion{\thanks{The full version of the paper can be accessed at \protect\url{https://arxiv.org/abs/1902.09310}}} % Replace URL with link to full paper or comment out this line

\title{\Large Accurately Estimating Unreported Infections using Information Theory}
\author{Jiaming Cui\thanks{College of Computing, Georgia Institute of Technology, Email: \{jiamingcui1997, badityap\}@gatech.edu.}$~$\thanks{Department of Computer Science, University of Virginia, Email: \{zej9va,ah8zf,ah3wj,vsakumar\}@virginia.edu.}$~$\thanks{Department of Computer Science, Virginia Tech, Email: jiamingcui@vt.edu.} 
\and Bijaya Adhikari\thanks{Department of Computer Science, The University of Iowa, Email: bijaya-adhikari@uiowa.edu.} 
\and Arash Haddadan$^\dagger$\thanks{Modeling and Optimization, Amazon, Email: ahaddada@amazon.com}
\and A S M Ahsan-Ul Haque$^\dagger$ 
\and Jilles Vreeken\thanks{CISPA Helmholtz Center for Information Security, Email: vreeken@cispa.de} 
\and Anil Vullikanti$^\dagger$
\and B. Aditya Prakash$^*$}
\date{}

\maketitle

% Copyright Statement
% When submitting your final paper to a SIAM proceedings, it is requested that you include
% the appropriate copyright in the footer of the paper.  The copyright added should be
% consistent with the copyright selected on the copyright form submitted with the paper.
% Please note that "20XX" should be changed to the year of the meeting.

% Default Copyright Statement
\fancyfoot[R]{\scriptsize{Copyright \textcopyright\ 2025 by SIAM\\
Unauthorized reproduction of this article is prohibited}}

% Depending on which copyright you agree to when you sign the copyright form, the copyright
% can be changed to one of the following after commenting out the default copyright statement
% above.

%\fancyfoot[R]{\scriptsize{Copyright \textcopyright\ 20XX\\
%Copyright for this paper is retained by authors}}

%\fancyfoot[R]{\scriptsize{Copyright \textcopyright\ 20XX\\
%Copyright retained by principal author's organization}}

%\pagenumbering{arabic}
%\setcounter{page}{1}%Leave this line commented out.

\begin{abstract}
{\small\baselineskip=9pt 
One of the most significant challenges in combating against the spread of infectious diseases was the difficulty in estimating the true magnitude of infections. Unreported infections could drive up disease spread, making it very hard to accurately estimate the infectivity of the pathogen, therewith hampering our ability to react effectively. Despite the use of surveillance-based methods such as serological studies, identifying the true magnitude is still challenging. This paper proposes an information theoretic approach for accurately estimating the number of total infections. Our approach is built on top of Ordinary Differential Equations (ODE) based models, which are commonly used in epidemiology and for estimating such infections. We show how we can help such models to better compute the number of total infections and identify the parametrization by which we need the fewest bits to describe the observed dynamics of reported infections. Our experiments on COVID-19 spread show that our approach leads to not only substantially better estimates of the number of total infections but also better forecasts of infections than standard model calibration based methods. We additionally show how our learned parametrization helps in modeling more accurate what-if scenarios with non-pharmaceutical interventions. %Our results support earlier findings that most COVID-19 infections were unreported and non-pharmaceutical interventions indeed helped to mitigate the spread of the outbreak.
Our approach provides a general method for improving epidemic modeling which is applicable broadly. 
}
\end{abstract}


\section{Introduction}

One of the most significant challenges in combating against the spread of infectious diseases in population is estimating the number of total infections. Our inability in estimating unreported infections allows them to drive up disease transmission. For example, in the COVID-19 pandemic, a significant number of COVID-19 infections were unreported, due to various factors such as the lack of testing and asymptomatic infections~\cite{bai2020presumed,aguilar2020investigating,stockmaier2021infectious,shaman2020estimation,li2020substantial}. %Phylogenetic studies revealed that COVID-19 had locally spread in Washington state before early 2020, when active community surveillance was implemented~\cite{bedford2020cryptic}. 
There were only 23 reported infections in five major U.S. cities by March 1, 2020, but it has been estimated that there were in fact more than 28,000 total infections by then~\cite{NYTimes}, and spread the COVID-19 to the whole US. Similar trends were observed in other countries, such as in Italy, Germany, and the UK~\cite{tiwari2021data}. %Despite having more advanced surveillance techniques such as serological studies, estimating the total number of infections continues to be a challenge for COVID-19 response even today~\cite{reportproblem,irons2021estimating}.

\begin{figure*}[ht!]
\centering
\vspace{-2em}
\includegraphics[width=\textwidth]{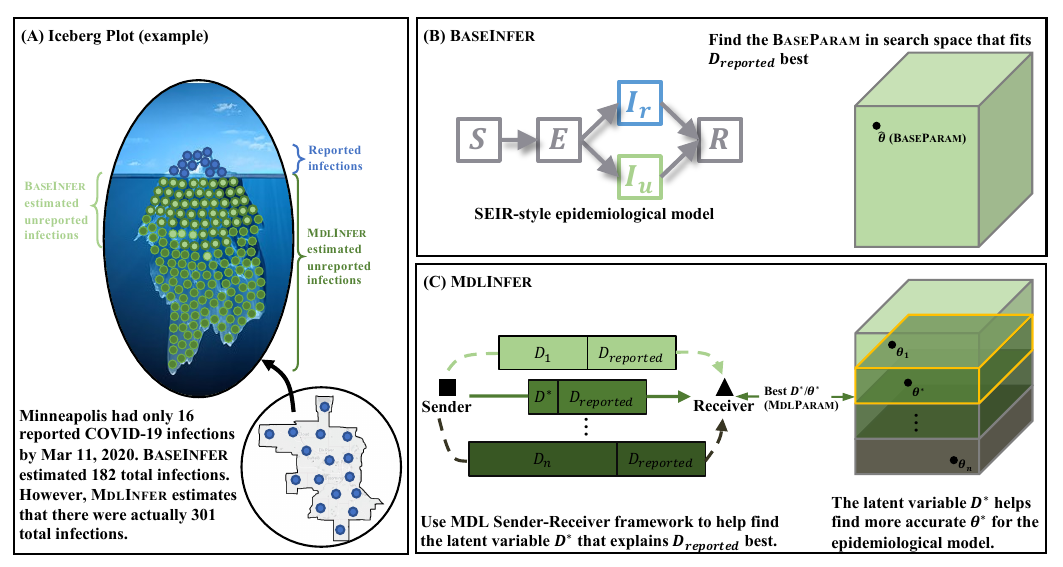}
\vspace{-2em}
\caption{\textbf{Overview of our problem and methodology.} (\textbf{A}) We visualize the idea of reported rates using the iceberg. The visible portion above water are the reported infections, which is only a fraction of the whole iceberg representing total infections. 
Light green corresponds to the 182 unreported infections estimated by typical current practice used by researchers. We call it as the basic approach, or $\BaseInfer$. In contrast, dark green corresponds to the more accurate and much larger 301 unreported infections found by our approach $\MDLFramework$. (\textbf{B}) The usual practice is to calibrate an epidemiological model to reported data and compute the reported rate from the resultant parameterization of the model. Here, an SEIR-style model with explicit compartments for reported-vs-unreported infection is shown in the figure as an example. (\textbf{C}) Our new approach $\MDLFramework$ instead aims to compute a more accurate reported rate by finding a `best' parametrization \textit{for the same epidemiological model} (i.e., SEIR-style model in this example) using a principled information theoretic formulation - two-part `sender-receiver' framework. Assume that a hypothetical Sender \emph{S} wants to transmit the reported infections as the $\textsc{Data}$  to a Receiver \emph{R} in the cheapest way possible. Hence \emph{S} will find/solve for the best $D^*$, intuitively, the $\textsc{Model}$ that takes the fewest number of bits to encode the $\textsc{Data}$. Using $D^*$, we can find the best $\Theta^*$ by exploring a smaller search space.}
\vspace{-1.5em}
\label{fig:Figure1}
\end{figure*}

In fact, an accurate estimation of the number of total infections is a fundamental epidemiological question and critical for pandemic planning and response. %Not withstanding its importance, there is not even a commonly agreed upon metric. One  proposal is the \emph{case ascertainment rate}, which is defined as the ratio of reported symptomatic infections to the actual number of symptomatic infections~\cite{Russell2020.07.07.20148460}. Another popular proposal is the 
Therefore, epidemiologists use \emph{reported rate} ($\alpha_{\mathrm{reported}}$) to capture total infections, which is defined as the ratio of reported infections to total infections~\cite{pei2021burden}. One of the benefits of using this definition is that it includes asymptomatic infections, which may also contribute substantially to spread~\cite{subramanian2021quantifying,moghadas2020implications}. %In this paper, we also focus on this particular measure. 
To estimate the reported rate, data scientists and epidemiologists have devoted much time and effort to using epidemiological models. %By now, 
There are many carefully constructed Ordinary Differential Equation (ODE) based models that capture the transmission dynamics of different infectious diseases~\cite{li2020substantial,Russell2020.07.07.20148460, Angelopoulos2020On,press2020modeling,kraemer2020effect,lai2020effect,wells2020impact,hao2020reconstruction,kain2021chopping,wilder2020modeling,wu2020substantial,cao2022micro,padmanabhan2022modeling}. However, these models still suffer from estimating accurate reported rates, leading to suboptimal total infections estimation. For example, as shown in Figure~\ref{fig:Figure1}, the Minneapolis Metro Area had only 16 COVID-19 reported infections by March 11, 2020. Although epidemiologists estimate that there were 182 total infections (light green part in the iceberg) using epi models, later studies revealed that there were actually around 300 total infections \textcolor{black}{(iceberg below the sea level)}~\cite{havers2020seroprevalence,CDCTracker}, which is much larger than  the epi model estimated values.

To tackle this, we propose a new information theory-based approach named $\MDLFramework$ to estimate the reported rate. It is based on the following central intuition: Suppose an ``oracle'' gives us the time series of the number of \emph{total infections} $D$, \textcolor{black}{we should be able to describe $D_{\mathrm{reported}}$ in a succinct way}: As we know $D$, it is trivial to get the reported rate $\alpha_{\mathrm{reported}}$. Then with both $D$ and $\alpha_{\mathrm{reported}}$, it will be trivial to describe $D_{\mathrm{reported}}$, as it is simply $D \times \alpha_{\mathrm{reported}}$ plus a little bit of noise. 

In practice, we are of course not given $D$, but we could estimate $D$ as a latent variable. Specifically, as shown in Figure~\ref{fig:Figure1}(C), we use Minimum Description Length (MDL) principle to estimate $D$, which allows us to most succinctly describe (i.e., most accurately encode/reconstruct) the dynamics of $D_{\mathrm{reported}}$. Here, $\MDLFramework$ gives an estimate of 301 total infections in Minneapolis as shown below the sea level, which is much closer to the ground truth total infections numbers of around 300~\cite{havers2020seroprevalence,CDCTracker}.

Our main contributions are summarized below.

\vspace{-0.25em}

\begin{itemize}
    \item We propose an MDL-based approach on top of ODE-based epidemiological models, which are harder to formulate and optimize. To the best of our knowledge, we are the first to propose an MDL-based approach on top of ODE-based epidemiological models.
    \vspace{-0.5em}
    \item Our proposed MDL-based approach $\MDLFramework$ performs superior to the state of the art epidemiological model in estimating total infections and predicting the future reported infections.
    \vspace{-0.5em}
    \item We also show that $\MDLFramework$ can aid policy making by analyzing counter-factual non-pharmaceutical interventions, while inaccurate epidemiological model estimates may lead to wrong non-pharmaceutical intervention conclusions.
\end{itemize}
The rest of the paper is organized in the following way: Section~\ref{sec:sec2} discusses the related works. In section~\ref{sec:sec3}, we introduce the current ODE model calibration method to estimate the reported rate, and the background of MDL framework. We then introduce our $\MDLFramework$ framework in section~\ref{sec:sec4} and explain how we use it to estimate the total infections. In section~\ref{sec:sec5}, we evaluate the performance of $\MDLFramework$. We then discuss future work and conclude in section~\ref{sec:sec6}.

\vspace{-0.5em}

\section{Related work}\label{sec:sec2}

\subsection{Reported rate estimation}

One of the most effective current methods to identify the reported rate in a region is through large-scale serological studies~\cite{sood2020seroprevalence, havers2020seroprevalence, zhang2020analysis}. These surveys use blood tests to identify the prevalence of antibodies against target pandemic in a large population. %For example, during the COVID-19 pandemic, the CDC COVID Data Tracker portal~\cite{CDCTracker, havers2020seroprevalence} summarizes the results of serological studies conducted by commercial laboratories at a national level as well as at 10 specific sites. %For example, the estimated reported rate was at most 0.1 in Minneapolis and South Florida as of April 2020. This means that there were at least 10 times more total infections than reported infections. 
While serological studies can give an accurate estimation, they are expensive and are not sustainable in the long run~\cite{testprice}. Furthermore, it is also challenging to obtain real-time data using such studies since there are unavoidable delays between sample collection and laboratory tests~\cite{CDCTracker, havers2020seroprevalence}. %Additional difficulties include sampling biases that make it necessary to use carefully designed heuristics to account for them~\cite{accorsi2021detect}. %Other methods include exploiting existing surveillance systems of related diseases like influenza, and using them to estimate symptomatic infections~\cite{lu2021estimating}. However, this can also be unreliable and requires ad-hoc corrections to account for the similarities between COVID-19 and influenza symptoms.

\vspace{-0.21em}
\subsection{Minimum Description Length framework}

MDL frameworks has been widely used for numerous optimization problems ranging from network summarization~\cite{koutra2014vog}, causality inference~\cite{budhathoki2018origo}, and failure detection in critical infrastructures~\cite{adhikari2018near}. \textcolor{black}{
They are also used in machine learning as regularization to help with model selection and avoid overfitting.~\cite{blier2018description}} However, these works are built on networks and agent-based models. To the best of our knowledge, we are the first to propose an MDL-based approach on top of ODE-based epidemiological models.

\vspace{-0.5em}

\section{Preliminaries}\label{sec:sec3}

\subsection{ODE-based Models}

An ODE-based epidemiological model uses ordinary differential equations to describe the spread of diseases by modeling changes in populations (e.g., susceptible, infected, recovered) over time~\cite{jang2019}. \textcolor{black}{In general, the $O_{\mathrm{M}}$ has a set of parameters $\Theta$ that need to estimate from \emph{observed data} using a so-called calibration procedure, $\Calibrate$}. In practice, the data we use for calibration can be the time series of the number of reported infections, or $D_{\mathrm{reported}}$. To estimate the number of total infections, these models often explicitly include reported rate as one of their parameters, or include multiple parameters that jointly account for it. We call it as $\BaseInfer$ in later sections for brief. There are many calibration procedures commonly used in literature, such as RMSE-based~\cite{gopalakrishnan2021globally} or Bayesian approaches~\cite{kain2021chopping,hao2020reconstruction}. $\BaseInfer$ is generally a complex, high-demensional problem, since there are multiple parameters interacting with each other. To make matters worse, there exist many possible parametrizations that show similar performance (e.g. in RMSE, likelihood) yet correspond to vastly different reported rates, and $\BaseInfer$ cannot select between these competing parametrizations in a principled way.%: the parametrization $\hat{\Theta}$ it results in may or may not overfit the reported infections and may or may not predict future infections well. One method for selecting them is to take a Bayesian approach. That is, we choose a prior distribution, and then select the best parametrization that maximizes the posterior probability. Choosing such a prior, however, is ad-hoc and does not generalize well across different models $O_{\mathrm{M}}$. As we will see in the experimental evaluation, minor differences in estimates of reported rates can indeed lead to very different forecasts of future trends and therewith intervention policy recommendations. 

%We call the above general methodology the basic approach to estimate the reported rate, or $\BaseInfer$ for short. It takes the epidemiological model $O_{\mathrm{M}}$, a calibration procedure $\Calibrate$, and observed data $D_{\mathrm{reported}}$ as input. The output of $\BaseInfer$ is then a baseline parametrization $\hat{\Theta}$ and, by extension, an estimated reported rate $\hat{\alpha}_{\mathrm{reported}}$. 

\vspace{-0.5em}
\subsection{Two-part sender-receiver MDL framework}

In this work, we use this framework to identify the total infections. The conceptual goal of the framework is to transmit the \textsc{Data} from the possession of the hypothetical sender \emph{S} to the hypothetical receiver \emph{R}. We assume the sender does this by first sending a \textsc{Model} and then sending the \textsc{Data} under this \textsc{Model}. In this MDL framework, we want to minimize the number of bits for this process. We do this by identifying the \textsc{Model} that encodes the \textsc{Data} such that the total number of bits needed to encode both the \textsc{Model} and the \textsc{Data} is minimized. \textcolor{black}{Hence, our cost function in the total number of bits needed is composed of two parts:} (i) model cost $L(\textsc{Model})$: The cost in bits of encoding the $\textsc{Model}$ and (ii) data cost $L(\textsc{Data}|\textsc{Model})$: The cost in bits of encoding the $\textsc{Data}$ given the $\textsc{Model}$. Intuitively, the idea is that a good $\textsc{Model}$ will lead to a fewer number of bits needed to encode both $\textsc{Model}$ and $\textsc{Data}$. The general MDL optimization problem can be formulated as follows: Given the $\textsc{Data}$, $L(\textsc{Model})$, and $L(\textsc{Data}|\textsc{Model})$, find $\textsc{Model}^{*}$ such that

\vspace{-1em}
\begin{equation*}
    \textsc{Model}^{*} = \arg \min_{\textsc{Model}} L(\textsc{Model})+L(\textsc{Data}|\textsc{Model})
    \label{eq:MDL0}
\end{equation*}
\vspace{-2em}

\vspace{-0.5em}
\begin{table}
    \centering
  \caption{List of notations}
  \vspace{-1em}
  \label{table:graph_notation}
  \scalebox{0.85}{
  \begin{tabular}{c|c}
    \toprule
    Notation & Description\\
    \midrule
    $O_{\mathrm{M}}$ & ODE model \\
    $\Theta$ & ODE model parameters to infer \\
    $D_{\mathrm{reported}}$ & Reported infections\\
    $\alpha_{\mathrm{reported}}$ & Reported rate \\
    $\BaseInfer$ & Baseline ODE calibration procedure \\
    & (calibrated on only $D_{\mathrm{reported}}$) \\ 
    $\hat{\Theta}$ & Parametrization estimated by $\BaseInfer$  \\
    $\hat{\alpha}_{\mathrm{reported}}$ & Reported rate in $\hat{\Theta}$ \\
    $D_{\mathrm{reported}}(\hat{\Theta})$ & ODE simulated reported infections using $\hat{\Theta}$ \\
    $D(\hat{\Theta})$ & ODE simulated total infections using $\hat{\Theta}$ \\
    $\MDLFramework$ & Our framework \\
    $D$ & Candidate total infections in $\MDLFramework$ \\
    $\Theta^{'}$ & Parametrization estimated by $\MDLFramework$ \\
    & when calibrating on both $D$ and $D_{\mathrm{reported}}$ \\
    $\alpha_{\mathrm{reported}}'$ & Reported rate in $\Theta'$ \\
    $D_{\mathrm{reported}}(\Theta')$ & ODE simulated reported infections using $\Theta'$ \\
    $D(\Theta')$ & ODE simulated total infections using $\Theta'$ \\
   \bottomrule
\end{tabular}
}
\vspace{-1.5em}
\end{table}

\section{\MDLFramework}\label{sec:sec4}

\subsection{MDL formulation}

In our situation, the $\textsc{Data}$ is the reported infections $D_{\mathrm{reported}}$, which is the only real-world data given to us. %Note that total infections are not directly observed. 
As for the $\textsc{Model}$, intuitively it should be $(D,\alpha_{\mathrm{reported}}')$ \textcolor{black}{since our goal is to find the total infections $D$ with the corresponding reported rate $\alpha_{\mathrm{reported}}'$.} Note that as two-part MDL (and MDL in general) does not assume the nature of the $\textsc{Data}$ or the $\textsc{Model}$, our $\MDLFramework$ can be applied to any ODE model. Next, we give more details how to formulate our problem of estimating total infections $D$. We also list the notations in Table~\ref{table:graph_notation}

%First, we need to introduce some notations. Given an epidemiological model $O_{\mathrm{M}}$ and the paramterization $\hat{\Theta}$ estimated by $\BaseInfer$, we can compute the reported infections. However, this is only an estimate of the reported infections rather than the exact $D_{\mathrm{reported}}$. This is because even though we have already calibrated $O_{\mathrm{M}}$ using $D_{\mathrm{reported}}$, the calibration cannot be perfect, and there will be differences between these estimated reported infections and $D_{\mathrm{reported}}$. Here, we term this estimated reported infections as $D_{\mathrm{reported}}(\hat{\Theta})$. We can also estimate the total infections $D(\hat{\Theta})$ for $O_{\mathrm{M}}$ in the same way. Similarly, we have the $D_{\mathrm{reported}}(\Theta^{'})$ and $D(\Theta^{'})$ for $\Theta^{'}$. As described in the introduction section, we can also calculate the reported rate $\hat{\alpha}_{\mathrm{reported}}$ and $\alpha_{\mathrm{reported}}'$ using $\hat{\Theta}$ and $\Theta^{'}$. With these notations, next we will formulate the space of all possible $\textsc{Model}$s and give the equation for the cost in bits of encoding $\textsc{Model}$ and $\textsc{Data}$. 

\subsubsection{$\textsc{Model}$ space}

As described above, our $\textsc{Model}$ is intuitively $(D,\alpha_{\mathrm{reported}}')$. Note that reported rate is actually one of the parameters for the ODE model $O_{\mathrm{M}}$, we choose to include its corresponding parametrization $\Theta^{'}$ into $\textsc{Model}$. \textcolor{black}{We further choose to add $\hat{\Theta}$ estimated by $\BaseInfer$, making our $\textsc{Model}$ to be $(D, \Theta^{'}, \hat{\Theta})$}. With $\textsc{Model}=(D, \Theta^{'}, \hat{\Theta})$, our $\textsc{Model}$ space will be all possible daily sequences for $D$ and all possible parametrizations for $\Theta^{'}$ and $\hat{\Theta}$. The MDL framework will search in this space to find the $\textsc{Model}^{*}$. We also discuss other alternative $\textsc{Model}$s and why $(D, \Theta^{'}, \hat{\Theta})$ is better in Appendix. 

\vspace{-0.25em}

\subsubsection{Model cost}

With $\textsc{Model} = (D, \Theta^{'}, \hat{\Theta})$, we conceptualize the model cost by imagining that the sender \emph{S} will send the $\textsc{Model} = (D, \Theta^{'}, \hat{\Theta})$ to the receiver \emph{R} in three parts: (i) first send the $\hat{\Theta}$ by encoding $\hat{\Theta}$ directly (ii) next send the $\Theta^{'}$ given $\hat{\Theta}$ by encoding $\Theta^{'}-\hat{\Theta}$ and (iii) then send $D$ given $\Theta^{'}$ and $\hat{\Theta}$ by encoding $\alpha_{\mathrm{reported}}' \times D-D_{\mathrm{reported}}(\hat{\Theta})$. Intuitively, both $\alpha_{\mathrm{reported}}' \times D$ and $D_{\mathrm{reported}}(\hat{\Theta})$ should be close to $D_{\mathrm{reported}}$, and the receiver could recover the $D$ using $\hat{\Theta}$, $\alpha_{\mathrm{reported}}'$, and $D_{\mathrm{reported}}(\hat{\Theta})$ as they have already been sent. We term the model cost as $L(D,\Theta^{'},\hat{\Theta})$ with three components: $\Cost(\hat{\Theta})$, $\Cost(\Theta^{'}|\hat{\Theta})$, and $\Cost(D|\Theta^{'},\hat{\Theta})$. Hence
\begin{equation*}
\scalebox{0.8}{
    $\begin{aligned}
        L(D,\Theta^{'},\hat{\Theta}) &= \Cost(\hat{\Theta}) + \Cost(\Theta^{'}-\hat{\Theta}|\hat{\Theta}) 
        \\ &+ \Cost(\alpha_{\mathrm{reported}}' \times D-D_{\mathrm{reported}}(\hat{\Theta})|\Theta^{'},\hat{\Theta})
    \end{aligned}$
    }
    \label{eq:ModelCost2}
\end{equation*}

Here, the $\Cost(\cdot)$ function gives the total number of bits we need to spend in encoding each term. The details of the encoding method can be found in the Appendix.

\vspace{-0.5em}

\subsubsection{Data cost}

We need to send the $\textsc{Data}=D_{\mathrm{reported}}$ next given the $\textsc{Model}$. Given $\textsc{Model} = (D, \Theta^{'}, \hat{\Theta})$, we send $\textsc{Data}$ by encoding $\frac{D-D_{\mathrm{reported}}}{1-\alpha_{\mathrm{reported}}'}-D(\Theta^{'})$. Intuitively, $D-D_{\mathrm{reported}}$ corresponds to the unreported infections, and $1-\alpha_{\mathrm{reported}}'$ is the unreported rate. Therefore, $\frac{D-D_{\mathrm{reported}}}{1-\alpha_{\mathrm{reported}}'}$ should be close to the total infections $D$ and $D(\Theta^{'})$. The receiver could also recover the $D_{\mathrm{reported}}$ using $D$, $\alpha_{\mathrm{reported}}'$, and $D(\Theta^{'})$ as they have already been sent. We term data cost as $L(D_{\mathrm{reported}}|D,\Theta^{'},\hat{\Theta})$ and formulate it as follows
\begin{equation*}
\scalebox{0.8}{
    $\begin{aligned}
        L(D_{\mathrm{reported}}|D,\Theta^{'},\hat{\Theta}) = \Cost(\frac{D-D_{\mathrm{reported}}}{1-\alpha_{\mathrm{reported}}'}-D(\Theta^{'})|D,\Theta^{'},\hat{\Theta})
    \end{aligned}$}
    \vspace{-0.5em}
    \label{eq:DataCost2}
\end{equation*}

\subsubsection{Total cost}
With  $L(D,\Theta^{'},\hat{\Theta})$ and $L(D_{\mathrm{reported}}|D,\Theta^{'},\hat{\Theta})$ above, the total cost  $L(D_{\mathrm{reported}},D,\Theta^{'},\hat{\Theta})$ will be:

\begin{equation*}
    \scalebox{0.8}{
    \vspace{-1em}
    $\begin{aligned}
        L(D_{\mathrm{reported}},D,\Theta^{'},\hat{\Theta}) &= L(D,\Theta^{'},\hat{\Theta}) + L(D_{\mathrm{reported}}|D,\Theta^{'},\hat{\Theta}) \\
        &= \Cost(\hat{\Theta}) + \Cost(\Theta^{'}-\hat{\Theta}|\hat{\Theta}) \\ 
        &+ \Cost(\alpha_{\mathrm{reported}}' \times D-D_{\mathrm{reported}}(\hat{\Theta})|\Theta^{'},\hat{\Theta}) \\
        &+ \Cost(\frac{D-D_{\mathrm{reported}}}{1-\alpha_{\mathrm{reported}}'}-D(\Theta^{'})|D,\Theta^{'},\hat{\Theta})    
    \end{aligned}$}
    \label{eq:MDL2}
\end{equation*}

\vspace{-0.5em}
\subsection{Problem statement}\label{sec:sec42}
Note that our main objective is to estimate the total infections $D$. With $L(D_{\mathrm{reported}},D,\Theta^{'},\hat{\Theta})$, we can state the problem as: Given the time sequence $D_{\mathrm{reported}}$, epidemiological model $O_{\mathrm{M}}$, and a calibration procedure $\Calibrate$, find $D^{*}$ that minimizes the MDL total cost i.e. 
\begin{equation*}
    D^{*} = \arg \min_{D} ~L(D_{\mathrm{reported}},D,\Theta^{'},\hat{\Theta})
    \label{eq:MDL6}
\end{equation*}
\vspace{-1.5em}

\vspace{-0.5em}
\subsection{Algorithm}
Next, we will present our algorithm to solve the problem in section~\ref{sec:sec42}. Note that directly searching $D^{*}$ naively is intractable since $D^{*}$ is a daily sequence not a scalar. Instead, we propose first finding a ``good enough'' reported rate $\alpha_{\mathrm{reported}}^{*}$ quickly with the constraint $D = \frac{D_{\mathrm{reported}}}{\alpha_{\mathrm{reported}}^{*}}$ to reduce the search space.
Then with this $\alpha_{\mathrm{reported}}^{*}$, we can search for the optimal $D^{*}$. Hence we propose a two-step algorithm: (i) do a linear search to find a good reported rate $\alpha_{\mathrm{reported}}^{*}$ (ii) given the $\alpha_{\mathrm{reported}}^{*}$ found above, use an optimization method to find the $D^{*}$ that minimizes $L(D_{\mathrm{reported}},D,\Theta^{'},\hat{\Theta})$ with $\alpha_{\mathrm{reported}}^{*}$ constraints. The pseudo-code is given in Algorithm~\ref{code}.

\begin{algorithm}[H]
  \caption{$\MDLFramework$}
  \label{code}
  \begin{algorithmic}[1]
  \REQUIRE $O_{\mathrm{M}}$, Calibration procedure $\Calibrate$, and $D_{\mathrm{reported}}$ 
  \STATE Calibrate $\hat{\Theta} = \Calibrate(O_{\mathrm{M}},D_{\mathrm{reported}})$
  \STATE The array to save the MDL cost: CostArray = [$~$]
  \FOR{$\alpha_{\mathrm{reported}}$ in the grid search space from 0.01 to 1 with step 0.01}
  \STATE $D = \frac{D_{\mathrm{reported}}}{\alpha_{\mathrm{reported}}}$
  \STATE Calibrate $\Theta^{'} = \Calibrate(O_{\mathrm{M}},(D,D_{\mathrm{reported}}))$
  \STATE CostArray[$\alpha_{\mathrm{reported}}$] = $L(D_{\mathrm{reported}}, D, \Theta^{'}, \hat{\Theta})$
  \ENDFOR
  \STATE $\alpha_{\mathrm{reported}}^{*} = \arg \min_{\alpha_{\mathrm{reported}}} $ CostArray[$\alpha_{\mathrm{reported}}$]
    \STATE Find the $D^{*}= \arg \min_{D} L(D_{\mathrm{reported}}, D, \Theta^{'}, \hat{\Theta})$. (using the Nelder-Mead algorithm).
  \ENSURE Total infections $D^{*}$
  \end{algorithmic}
\end{algorithm}
\vspace{-1em}

\vspace{-0.5em}
\section{Experiments}\label{sec:sec5}

\begin{figure*}[t]
\centering
\vspace{-2em}
\includegraphics[width=\textwidth]{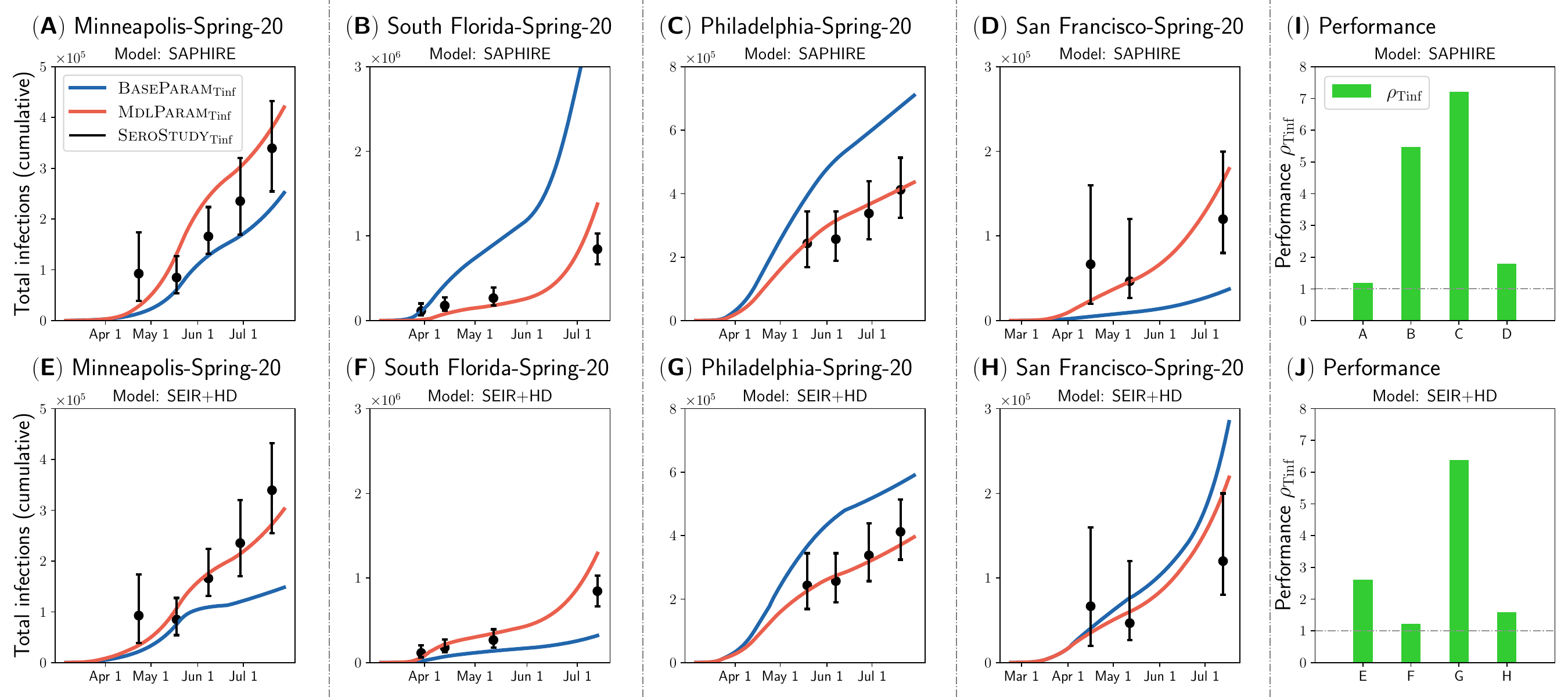}
\vspace{-2em}
\caption{\textbf{$\MDLFramework$ (red) gives a closer estimation of total infections to serological studies (black) than $\BaseInfer$ (blue) on various geographical regions and time periods.} Note that both approaches try to fit the serological studies without being informed with them. (\textbf{A})-(\textbf{H}) The red and blue curves represent $\MDLFramework$'s estimation of total infections, $\MDLResult_{\mathrm{Tinf}}$, and $\BaseInfer$'s estimation of total infections, $\MordecaiResult_{\mathrm{Tinf}}$, respectively. The black point estimates and confidence intervals represent the total infections estimated by serological studies~\cite{CDCTracker, havers2020seroprevalence}, $\textsc{SeroStudy}_{\mathrm{Tinf}}$. (\textbf{A})-(\textbf{D}) use $\SAPHIRE$ model and (\textbf{E})-(\textbf{H}) use $\SEIR$ model. (\textbf{I})-(\textbf{J}) The performance metric, $\rho_{\mathrm{Tinf}}$, comparing $\MDLResult_{\mathrm{Tinf}}$ against $\MordecaiResult_{\mathrm{Tinf}}$ in fitting serological studies is shown for each region. (\textbf{I}) is for $\SAPHIRE$ model in (\textbf{A})-(\textbf{D}), and (\textbf{J}) is for $\SEIR$ model in (\textbf{E})-(\textbf{H}). Here, the values of $\rho_{\mathrm{Tinf}}$ are 1.20, 5.47, 7.21, and 1.79 in (\textbf{I}), and 2.62 ,1.22, 6.39, and 1.58 in (\textbf{J}). Note that $\rho_{\mathrm{Tinf}}$ larger than 1 means that $\MDLResult_{\mathrm{Tinf}}$ is closer to $\textsc{SeroStudy}_{\mathrm{Tinf}}$ than $\MordecaiResult_{\mathrm{Tinf}}$. We show more experiments in the Appendix.}
\vspace{-1.5em}
\label{fig:Figure2}
\end{figure*}

In this section, we will answer the following research questions

\begin{itemize}
    \item \textbf{Question 1:} Can $\MDLFramework$ estimate the total infections accurately than $\BaseInfer$ and fit the large-scale serological studies~\cite{sood2020seroprevalence, havers2020seroprevalence, zhang2020analysis}?
    \vspace{-0.5em}
    \item \textbf{Question 2:} Can $\MDLFramework$ fit the reported infection $D_{\mathrm{reported}}$ and forecast future infections accurately?
    \vspace{-0.5em}
    \item \textbf{Question 3:} How does $\MDLFramework$ captures the trends of symptomatic rate?
    \vspace{-0.5em}
    \item \textcolor{black}{\textbf{Question 4:} How can $\MDLFramework$ help to evaluate the non-pharmaceutical interventions}?
\end{itemize}

%Next, we present our empirical findings on a large set of experiments in different geographical regions and time periods. 

\subsection{Setup}
\vspace{-0.75em}
\subsubsection{Dataset}
\vspace{-0.75em}
We choose 8 regions and periods based on the severity of the outbreak and the availability of serological studies and symptomatic surveillance data. The serological studies dataset consists of the point and 95\% confidence interval estimates of the prevalence of antibodies to SARS-CoV-2 in these locations every 3–4 weeks from March to July 2020~\cite{havers2020seroprevalence,CDCTracker}. The symptomatic surveillance dataset consists of point estimate $\textsc{Rate}_{\mathrm{Symp}}$ and standard error of the COVID-related symptomatic rate starting from April 6, 2020~\cite{delphisurvey,salomon2021us}. The reported infections are from New York Times~\cite{nytimeswebsite}, which consists of the daily time sequence of reported COVID-19 infections $D_{\mathrm{reported}}$ and the mortality $D_{\mathrm{mortality}}$ (cumulative values) for each county in the US starting from January 21, 2020. In each region, we divide the timeline into two time periods: (i) observed period, when only the number of reported infections are available, and both $\BaseInfer$ and $\MDLFramework$ are used to learn the baseline parametrization ($\MordecaiResult$) $\hat{\Theta}$ and MDL parametrization ($\MDLResult$) $\Theta^*$, and (ii) forecast period, where we evaluate the forecasts generated by the parametrizations learned in the observed period. To handle the time-varying reported rates, we divide the observed period into multiple sub-periods and learn different reported rates for each sub-period separately. Our data and code have been deposited in \hyperlink{https://github.com/AdityaLab/MDL-ODE-Missing}{https://github.com/AdityaLab/MDL-ODE-Missing}, which can be run on other datasets. A demo is also deposited there. 

\begin{figure*}[t]
\centering
\vspace{-2em}
\includegraphics[width=\textwidth]{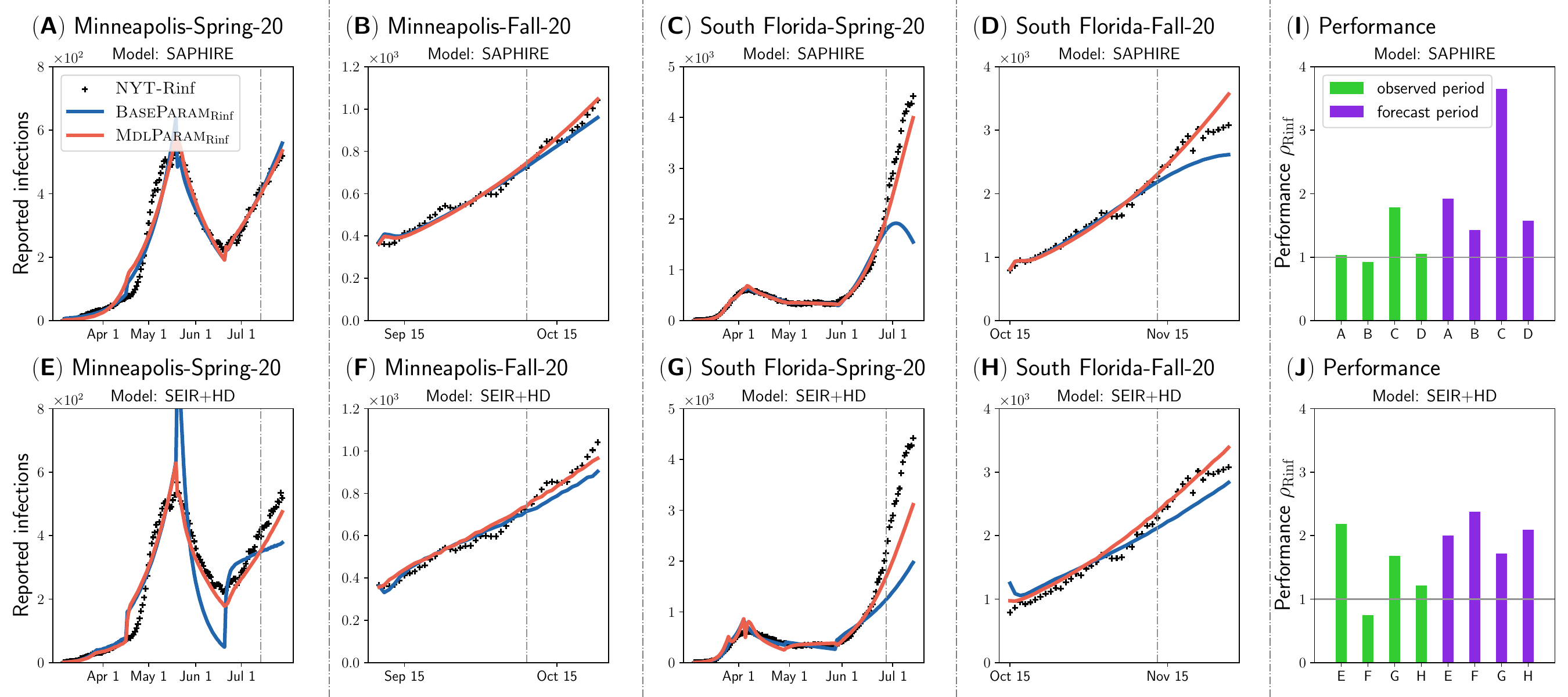}
\vspace{-2em}
\caption{\textbf{$\MDLFramework$ (red) gives a closer estimation of reported infections (black) than $\BaseInfer$ (blue) on various geographical regions and time periods.} We use the reported infections in the observed period as inputs and try to forecast the future reported infections (forecast period). (\textbf{A})-(\textbf{H}) The vertical grey dash line divides the observed period (left) and forecast period (right). The red and blue curves represent $\MDLFramework$'s estimation of reported infections, $\MDLResult_{\mathrm{Rinf}}$, and $\BaseInfer$'s estimation of reported infections, $\MordecaiResult_{\mathrm{Rinf}}$, respectively. The black plus symbols represent the reported infections collected by the New York Times ($\textsc{NYT-R}\mathrm{inf}$). (\textbf{A})-(\textbf{D}) use $\SAPHIRE$ model and (\textbf{E})-(\textbf{H}) use $\SEIR$ model. (\textbf{I})-(\textbf{J}) The performance metric, $\rho_{\mathrm{Rinf}}$, comparing $\MDLResult_{\mathrm{Rinf}}$ against $\MordecaiResult_{\mathrm{Rinf}}$ in fitting reported infections is shown for each region. (\textbf{I}) is for $\SAPHIRE$ model in (\textbf{A})-(\textbf{D}), and (\textbf{J}) is for $\SEIR$ model in (\textbf{E})-(\textbf{H}). Note that $\rho_{\mathrm{Rinf}}$ larger than 1 means that $\MDLResult_{\mathrm{Rinf}}$ is closer to $\textsc{NYT-R}\mathrm{inf}$ than $\MordecaiResult_{\mathrm{Rinf}}$. We show more experiments in the Appendix.}
\vspace{-1.5em}
\label{fig:Figure3}
\end{figure*}

\vspace{-0.5em}
\subsubsection{ODE model}

We compare $\MDLFramework$ and $\BaseInfer$ using two different ODE-based epidemiological models: $\SAPHIRE$~\cite{hao2020reconstruction} and $\SEIR$~\cite{kain2021chopping} as $O_{\mathrm{M}}$. Following their literature~\cite{hao2020reconstruction,kain2021chopping}, we use Markov Chain Monte Carlo ($\MCMC$) as the calibration procedure $\Calibrate$ for $\SAPHIRE$ and iterated filtering ($\IteratedFiltering$) for $\SEIR$, both of with are Bayesian approaches\cite{ionides2015inference}. Both these epidemiological models have previously been shown to perform well in fitting reported infections and provided insight that was beneficial for the COVID-19 response.

\subsubsection{Metrics}\vspace{-0.5em}

To quantify the performance gap between the two approaches, we use the root mean squared error (RMSE) following the previous work~\cite{rodriguez2023einns,rodriguez2021deepcovid,cramer2022evaluation,cramer2022united} for evaluation. To further demonstrate the performance, we further compute the ratio $\rho$ as the fraction of the RMSE errors of $\BaseInfer$ over $\MDLFramework$. Specifically, when the ratio is greater than 1, it implies that the $\MDLFramework$ is performing $\rho$ times better than $\BaseInfer$.

 %$\SAPHIRE$ focuses on two key features of the outbreak: high covertness and high transmissibility that drove the outbreak of COVID-19 in Wuhan. $\SEIR$ investigates how non-pharmaceutical interventions like social distancing will be needed to maintain epidemic control. These models are broadly representative to show that $\MDLFramework$ gives consistent performance across multiple epidemiological models with different dynamics. The experiments clearly show that our proposed MDL-based approach $\MDLFramework$ performs superior to the state of the art. To illustrate, we give an example in Figure~\ref{fig:Figure1}. By March 11, 2020, the Minneapolis Metro Area had only 16 COVID-19 reported infections. $\BaseInfer$ estimated 182 total infections, which are colored as light green in the iceberg. On the other hand, our $\MDLFramework$ gives an estimate of 301 total infections shown below the sea level, which is closer to the total infections estimated from serological studies~\cite{havers2020seroprevalence,CDCTracker}. Additionally, $\MDLFramework$ also leads to better fits and \emph{future} projections on reported infections. We also demonstrate that $\MDLFramework$ can aid policy making by analyzing counter-factual non-pharmaceutical interventions, while inaccurate $\BaseInfer$ estimates lead to wrong non-pharmaceutical intervention conclusions.

\vspace{-0.5em}
\subsection{Q1: Estimating total infections}~\label{sec:exp1}

\begin{figure*}[t]
\centering
\vspace{-2em}
\includegraphics[width=\textwidth]{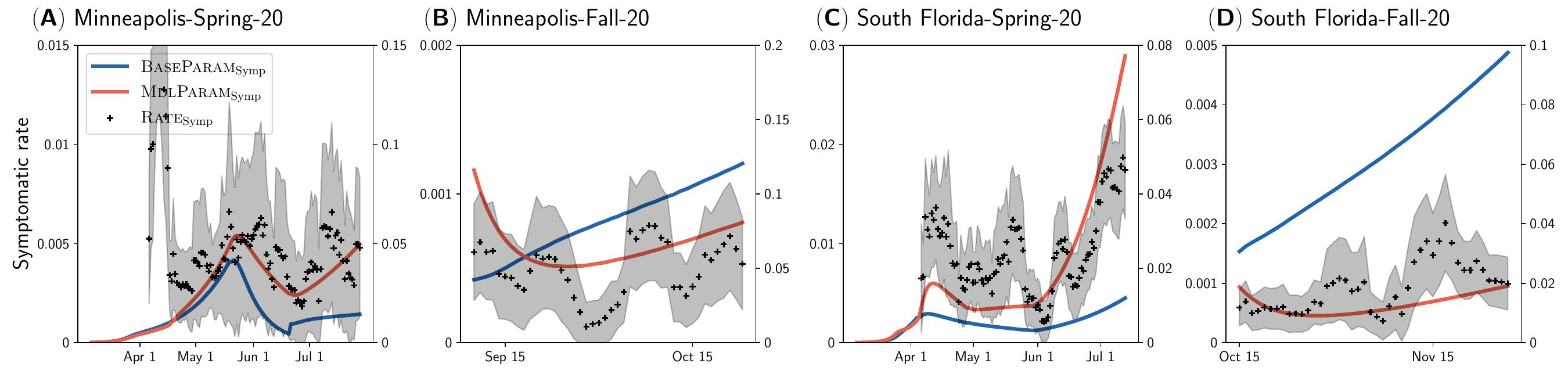}
\vspace{-2em}
\caption{\textbf{$\MDLFramework$ (red) gives a closer estimation of the trends of symptomatic rate (black) than $\BaseInfer$ (blue) on various geographical regions and time periods.} (\textbf{A})-(\textbf{D}) The red and blue curves represent $\MDLFramework$'s estimation of symptomatic rate, $\MDLResult_{\mathrm{Symp}}$, and $\BaseInfer$'s estimation of symptomatic rate, $\MordecaiResult_{\mathrm{Symp}}$, respectively. They use the y-scale on the left. The black points and the shaded regions are the point estimate with standard error for $\textsc{Rate}_{\mathrm{Symp}}$ (the COVID-related symptomatic rates derived from the symptomatic surveillance dataset~\cite{delphisurvey,salomon2021us}). They use the y-scale on the right. Note that we focus on trends instead of the exact numbers, hence $\MDLResult_{\mathrm{Symp}}$/$\MordecaiResult_{\mathrm{Symp}}$, and $\textsc{Rate}_{\mathrm{Symp}}$ may scale differently. We show more experiments in the Appendix.}
\label{fig:Figure4}
\vspace{-1.5em}
\end{figure*}

Here, we use the point estimates of the total infections calculated from serological studies as the ground truth (black dots shown in Figure~\ref{fig:Figure2}). We call it $\textsc{SeroStudy}_{\mathrm{Tinf}}$. We also plot $\MDLFramework$'s estimation of total infections, $\MDLResult_{\mathrm{Tinf}}$, in the same figure (red curve). To compare the performance of $\MDLFramework$ and $\BaseInfer$ with $\textsc{SeroStudy}_{\mathrm{Tinf}}$, we use the cumulative value of estimated total infections. Note that values from the serological studies are not directly comparable with the total infections because of the lag between antibodies becoming detectable and infections being reported~\cite{CDCTracker, havers2020seroprevalence}. In Figure~\ref{fig:Figure2}, we have already accounted for this lag following CDC study guidelines~\cite{CDCTracker, havers2020seroprevalence} (See Methods section for details). The vertical black lines shows a 95\% confidence interval for $\textsc{SeroStudy}_{\mathrm{Tinf}}$. The blue curve represents total infections estimated by $\BaseInfer$, $\MordecaiResult_{\mathrm{Tinf}}$. As seen in the figure, $\MDLResult_{\mathrm{Tinf}}$ falls within the confidence interval of the estimates given by serological studies. Significantly, in Figure~\ref{fig:Figure2}\textbf{B} and Figure~\ref{fig:Figure2}\textbf{F} for South Florida, $\BaseInfer$ for $\SAPHIRE$ model~\cite{hao2020reconstruction} overestimates the total infections, while for $\SEIR$ model underestimates the total infections. However, $\MDLFramework$ consistently estimates the total infections correctly. This observation shows that as needed, $\MDLResult_{\mathrm{Tinf}}$ can improve upon the $\MordecaiResult_{\mathrm{Tinf}}$ in either direction (i.e., by increasing or decreasing the total infections). Note that the $\MDLResult_{\mathrm{Tinf}}$ curves from both models are closer to the $\textsc{SeroStudy}_{\mathrm{Tinf}}$ even when the $\MordecaiResult_{\mathrm{Tinf}}$ curves are different. The results of better accuracy in spite of various geographical regions and time periods show that $\MDLFramework$ is consistently able to estimate total infections more accurately.

In Figure~\ref{fig:Figure2}\textbf{I} and Figure~\ref{fig:Figure2}\textbf{J}, we plot $\rho_{\mathrm{Tinf}}=\frac{\textsc{Rmse}(\MordecaiResult_{\mathrm{Tinf}},\textsc{SeroStudy}_{\mathrm{Tinf}})}{\textsc{Rmse}(\MDLResult_{\mathrm{Tinf}},\textsc{SeroStudy}_{\mathrm{Tinf}})}$. Overall, the $\rho_{\mathrm{Tinf}}$ values are greater than 1 in Figure~\ref{fig:Figure2}\textbf{I} and Figure~\ref{fig:Figure2}\textbf{J}, which indicates that $\MDLFramework$ performs better than $\BaseInfer$. Note that even when the value of $\rho_{\mathrm{Tinf}}$ is 1.20 for Figure~\ref{fig:Figure2}\textbf{A}, the improvement made by $\MDLResult_{\mathrm{Tinf}}$ over $\MordecaiResult_{\mathrm{Tinf}}$ in terms of RMSE is about 12091. Hence, one can conclude that $\MDLFramework$ is indeed superior to $\BaseInfer$, when it comes to estimating total infections. We show more experiments in the Appendix.

\vspace{-0.5em}
\subsection{Q2: Estimating reported infections} 

Here, we first use the observed period to learn the parametrizations. We then \emph{forecast} the future reported infections (i.e., forecast periods), which were \emph{not} accessible to the model while training. The results are summarized in Figure~\ref{fig:Figure3}. In Figure~\ref{fig:Figure3}\textbf{A} to Figure~\ref{fig:Figure3}\textbf{H}, the vertical grey dash line divides the observed and forecast period. The black plus symbols represent reported infections collected by the New York Times, $\textsc{NYT-R}\mathrm{inf}$. The red curve represents $\MDLFramework$'s estimation of reported infections, $\MDLResult_{\mathrm{Rinf}}$. Similarly, the blue curve represents $\BaseInfer$'s estimation of reported infections, $\MordecaiResult_{\mathrm{Rinf}}$. Note that the curves to the right of the vertical grey line are future predictions. As seen in Figure~\ref{fig:Figure3}, $\MDLResult_{\mathrm{Rinf}}$ aligns more closely with $\textsc{NYT-R}\mathrm{inf}$ than $\MordecaiResult_{\mathrm{Rinf}}$, indicating the superiority of $\MDLFramework$ in fitting and forecasting reported infections.

We use a similar performance metric $\rho_{\mathrm{Rinf}}=\frac{\textsc{Rmse}(\MordecaiResult_{\mathrm{Rinf}},\textsc{NYT-R}\mathrm{inf})}{\textsc{Rmse}(\MDLResult_{\mathrm{Rinf}},\textsc{NYT-R}\mathrm{inf})}$ to compare $\MDLResult_{\mathrm{Rinf}}$ against $\MordecaiResult_{\mathrm{Rinf}}$ in a manner similar to $\rho_{\mathrm{Tinf}}$. In Figure~\ref{fig:Figure3}\textbf{I} and Figure~\ref{fig:Figure3}\textbf{J}, we plot the $\rho_{\mathrm{Rinf}}$ for the observed and forecast period. In both periods, we notice that the $\rho_{\mathrm{Rinf}}$ is close to or greater than 1. This further shows that $\MDLFramework$ has a better or at least closer fit for reported infections than $\BaseInfer$. Additionally, the $\rho_{\mathrm{Rinf}}$ for the forecast period is even greater than $\rho_{\mathrm{Rinf}}$ for the observed period, which shows that $\MDLFramework$ performs even better than $\BaseInfer$ while forecasting.

Note that Figure~\ref{fig:Figure3}\textbf{A}, \textbf{C}, \textbf{E}, \textbf{G} correspond to the early state of the COVID-19 epidemic in spring and summer 2020, and Figure~\ref{fig:Figure3}\textbf{B}, \textbf{D}, \textbf{F}, \textbf{H} correspond to fall 2020. We can see that $\MDLFramework$ performs well in estimating temporal patterns at different stages of the COVID-19 epidemic. We show more experiments in the Appendix.

\vspace{-0.5em}
\subsection{Q3: Estimating symptomatic rate trends}~\label{sec:exp3}

We validate this observation using Facebook's symptomatic surveillance dataset~\cite{delphisurvey}. We plot $\MDLFramework$'s and $\BaseInfer$'s estimated symptomatic rate over time and overlay the estimates and standard error from the symptomatic surveillance data in Figure~\ref{fig:Figure4}. The red and blue curves are   $\MDLFramework$'s and $\BaseInfer$'s estimation of symptomatic rates, $\MDLResult_{\mathrm{Symp}}$ and $\MordecaiResult_{\mathrm{Symp}}$ respectively. Note that  $\SAPHIRE$ model does not contain states corresponding to the symptomatic infections. Therefore, we only focus on $\SEIR$ model. We compare the trends of the $\MDLResult_{\mathrm{Symp}}$ and $\MordecaiResult_{\mathrm{Symp}}$ with the symptomatic surveillance results. We focus on trends rather than actual values because the symptomatic rate numbers could be biased~\cite{delphisurvey} (see Methods section for a detailed discussion) and therefore cannot be compared directly with model outputs like what we have done for serological studies. As seen in Figure~\ref{fig:Figure4}, $\MDLResult_{\mathrm{Symp}}$ captures the trends of the surveyed symptomatic rate $\textsc{Rate}_{\mathrm{Symp}}$ (black plus symbols) better than $\MordecaiResult_{\mathrm{Symp}}$. We show more experiments in the Appendix.

To summarize, these three sets of experiments in section~\ref{sec:exp1} to section~\ref{sec:exp3} together demonstrate that $\BaseInfer$ fail to accurately estimate the total infections including unreported ones. On the other hand, $\MDLFramework$ estimates total infections closer to those estimated by serological studies and better fits reported infections and symptomatic rate trends.  

\vspace{-0.5em}
\subsection{Q4: Evaluate the effect of non-pharmaceutical Interventions}

\begin{figure}[t]
\centering
\includegraphics[width=0.5\textwidth]{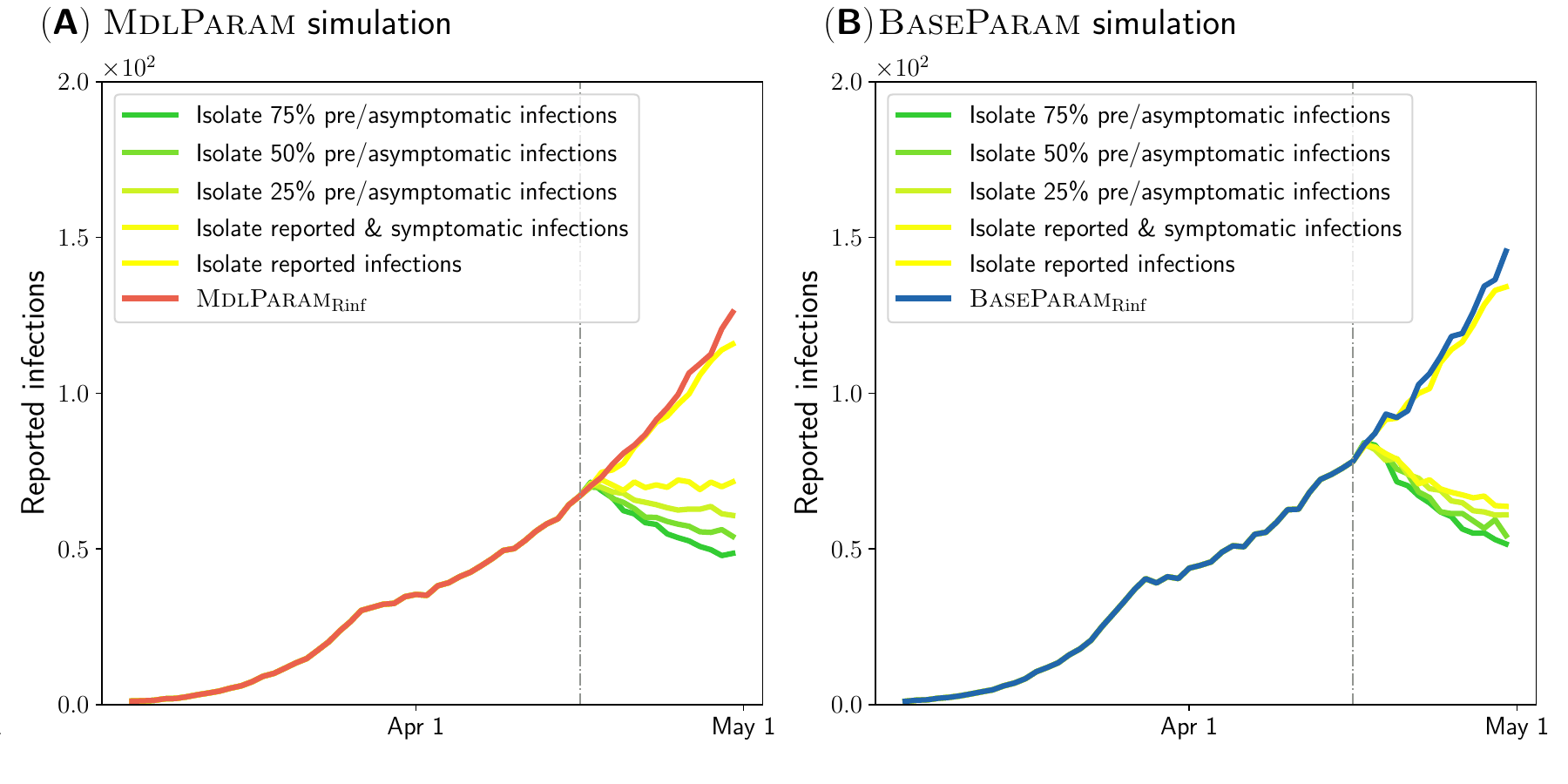}
%(\textbf{A}) $\MDLFramework$ estimates cumulative reported rate more accurately than $\BaseInfer$: The blue and red curve represent the cumulative reported rate estimated by $\BaseInfer$, $\MordecaiResult_{\mathrm{Rate}}$, and by $\MDLFramework$, $\MDLResult_{\mathrm{Rate}}$, respectively. The black point estimate and its confidence interval represent the cumulative reported rate $\textsc{SeroStudy}_{\mathrm{Rate}}$ estimated by serological studies~\cite{CDCTracker,havers2020seroprevalence}. Note that both approaches try to fit the $\textsc{SeroStudy}_{\mathrm{Rate}}$ without being informed with them. The results reveal that a large majority of COVID-19 infections were unreported. 
\vspace{-2.25em}
\caption{(\textbf{A}) $\MDLFramework$ reveals that non-pharmaceutical interventions (NPI) on asymptomatic and presymptomatic infections are essential to control the COVID-19 epidemic. Here, the red curve and other five curves represent the $\MDLFramework$'s estimation of reported infections for no NPI scenario and 5 different NPI scenarios described in the Results section. The vertical grey dash line divides the observed period (left) and forecast period (right). (\textbf{B}) 
Inaccurate estimation by $\BaseInfer$ may lead to wrong NPI conclusions. The blue curve and other five curves represent the $\BaseInfer$'s estimation of reported infections for no NPI scenario and the same 5 scenarios in (\textbf{B}).}
\vspace{-1.5em}
\label{fig:Figure5}
\end{figure}

We have already shown that $\MDLFramework$ is able to estimate the number of total infections accurately. In the following three observations, we show that such accurate estimations are important for evaluating the effect of non-pharmaceutical interventions.

%\subsubsection{$\MDLFramework$ reveals that a large majority of COVID-19 infections were unreported} 

%We compute the \emph{cumulative} reported rate $\MDLResult_{\mathrm{Rate}}$ measured by the ratio of the cumulative value of reported infections to the total infections estimated by $\MDLFramework$ over time and plotted it for Minneapolis-Spring-20 in Figure~\ref{fig:Figure5}\textbf{A}. The figure shows that the $\MDLResult_{\mathrm{Rate}}$ increases in early March, and then gradually decreases. This observation is explained by the community spread-driven COVID-19 outbreaks that were not reported until early March, which fits earlier studies~\cite{lu2021estimating}.

\vspace{-0.5em}
\subsubsection{Non-pharmaceutical interventions on asymptomatic and presymptomatic infections are essential to control the COVID-19 epidemic} Our simulations show that non-pharmaceutical interventions on asymptomatic and presymptomatic infections are essential to control COVID-19. Here, we plot the simulated reported infections of $\MDLResult$ in Figure~\ref{fig:Figure5}\textbf{A} (red curve). We then repeat the simulation of reported infections for 5 different scenarios: (i) isolate just the reported infections, (ii) isolate just the symptomatic infections, and isolate symptomatic infections in addition to (iii) 25\%, (iv) 50\%, and (v) 75\% of both asymptomatic and presymptomatic infections.  In our setup, we assume that the infectivity reduces by half when a person is isolated. As seen in Figure~\ref{fig:Figure5}\textbf{A}, when only the reported infections are isolated, there is almost no change in the ``future'' reported infections. However, when we isolate both the reported and symptomatic infections, the reported infections decreases significantly. Even here, the reported infections are still not in decreasing trend. On the other hand, non-pharmaceutical interventions for some fraction of asymptomatic and presymptomatic infections make reported infections decrease. Thus, we can conclude that NPIs on asymptomatic infections are essential in controlling the COVID-19 epidemic.\par

\vspace{-0.5em}
\subsubsection{Accuracy of non-pharmaceutical intervention simulations relies on the good estimation of parametrization} Next, we also plot the simulated reported infections generated by $\BaseInfer$ in Figure~\ref{fig:Figure5}\textbf{B} (blue curve). As seen in the figure, based on $\BaseInfer$, we can infer that only non-pharmaceutical interventions on symptomatic infections are enough to control the COVID-19 epidemic. However, this has been proven to be incorrect by prior studies and real-world observations~\cite{moghadas2020implications}. Therefore, we can conclude that the accuracy of non-pharmaceutical intervention simulation relies on the quality of the learned parametrization.\par

\vspace{-0.5em}
\section{Conclusion}\label{sec:sec6}

This study proposes $\MDLFramework$, a data-driven model selection approach that automatically estimates the number of total infections using epidemiological models. Our approach leverages the information theoretic Minimum Description Length (MDL) principle and addresses several gaps in current practice including the long-term infeasibility of serological studies~\cite{havers2020seroprevalence}, and ad-hoc assumptions in epidemiological models~\cite{kain2021chopping,li2020substantial,pei2020differential,hao2020reconstruction}. Overall, $\MDLFramework$ is a robust data-driven method to accurately estimate total infections, which will help data scientists, epidemiologists, and policy-makers to further improve existing ODE-based epidemiological models, make accurate forecasts, and combat future pandemics. More generally, $\MDLFramework$ opens up a new line of research in epidemic modeling using information theory.

\noindent \textbf{Acknowledgements:} This paper was partially supported by the NSF (Expeditions CCF-1918770 and CCF-1918656, CAREER IIS-2028586, RAPID IIS-2027862, Medium IIS-1955883, Medium IIS-2106961, Medium IIS-2403240, IIS-1931628, IIS-1955797, IIS-2027848, IIS-2331315), NIH 2R01GM109718, CDC MInD program U01CK000589, ORNL, Dolby faculty research award, and funds/computing resources from Georgia Tech and GTRI. B. A. was in part supported by the CDC MInD-Healthcare U01CK000531-Supplement. A.V.'s work is also supported in part by grants from the UVA Global Infectious Diseases Institute (GIDI).

\clearpage
\newpage

\bibliographystyle{abbrv}
%\bibliography{reference.bib}

\clearpage
\newpage

\appendix

\setcounter{figure}{0}
\renewcommand{\figurename}{Figure}
\renewcommand{\thefigure}{S\arabic{figure}}

\setcounter{table}{0}
\renewcommand{\tablename}{Table}
\renewcommand{\thetable}{S\arabic{table}}

\noindent \textbf{\large Appendix}

\section{Data}

\subsection{New York Times reported infections~\cite{nytimeswebsite}}
This dataset ($\textsc{NYT-R}\mathrm{inf}$) consists of the time sequence of reported infections $D_{\mathrm{reported}}$ and reported mortality $D_{\mathrm{mortality}}$ in each county across the U.S. since the beginning of the COVID-19 pandemic (January 21, 2020) to current. For each county, the $\textsc{NYT-R}\mathrm{inf}$ dataset provides the date, FIPS code, and the cumulative values of reported infections and mortality. Here, we use the averaged counts over 14 days to eliminate noise.

\subsection{Serological studies~\cite{havers2020seroprevalence,CDCTracker}}
This dataset consists of the point estimate and 95\% confidence interval of the prevalence of antibodies to SARS-CoV-2 in 10 US locations every 3-4 weeks during March to July 2020. The serological studies use the blood specimens collected from population. For each location, CDC collects 1800 samples approximately every 3-4 weeks. Using the prevalence of the antibodies and the population, we can compute the estimated total infections and 95\% confidence interval in the location. However, we cannot compare this number with the epidemiological model estimated total infection numbers directly as mentioned in the main article Methods section. We account for this problem by comparing the serological studies numbers with the estimated total infections of 7 days prior to the first day of specimen collection period (as suggested by the CDC serological studies work~\cite{havers2020seroprevalence}).

\subsection{Symptomatic surveillance~\cite{delphisurvey}}
This dataset comes from Facebook's symptomatic survey~\cite{delphisurvey}. %The daily number of participants in this survey is 55,000 on average, and the number of total participants is 16,398,000 (as of January 28, 2021) started from April 6, 2020. 
The survey started on April 6, 2020 to current.
As of January 28, 2021, there were a total of 16,398,000 participants, with the average daily participants number of 55,000.  
The survey asks a series of questions designed to help researchers understand the spread of COVID-19 and its effect on people in the United States. For the signal, they estimate the percentage of self-reported COVID-19 symptoms in population defined as fever along with either cough, shortness of breath, or difficulty breathing~\cite{delphisurvey}. The dataset also includes weighted version which accounts for the differences between Facebook users and the United States population. In the experiments, we contrast the symptomatic rate trends inferred by our approach against the weighted data from the survey.

\section{ODE Model}

\subsection{SAPHIRE Model}

\begin{figure}[h]
    \centering
    \includegraphics[width=0.5\textwidth]{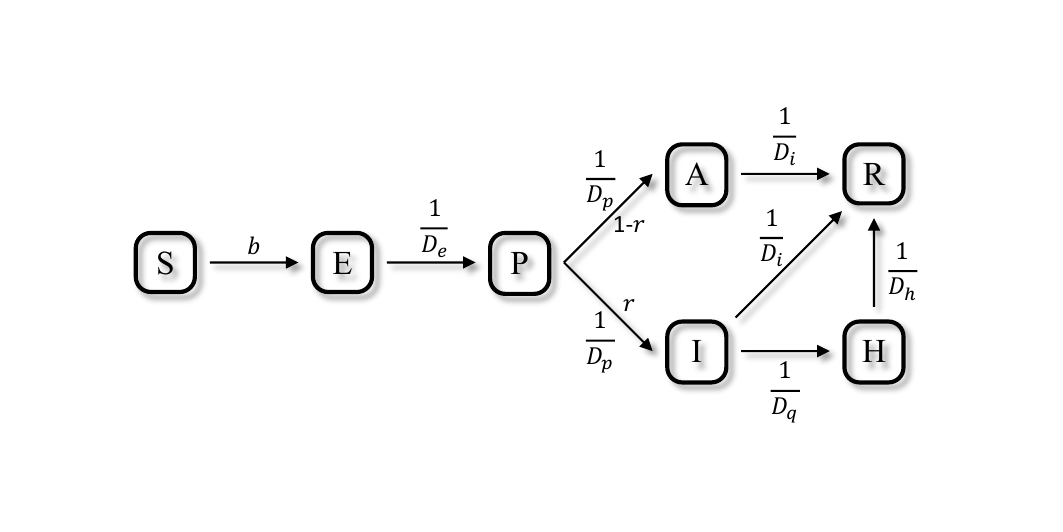}
    \caption{Compartmental diagram of SAPHIRE model~\cite{hao2020reconstruction}.}
    \label{fig:NatureDiagram}
\end{figure}

We use the SAPHIRE model~\cite{kain2021chopping} as one epidemiological model $O_{\mathrm{M}}$ in our experiments. The compartmental diagram of SAPHIRE model is shown in Figure~\ref{fig:NatureDiagram}. The SAPHIRE model has 9 different parameters. Note that only two parameters are calibrated, while the rest are fixed. 

In this article, we expect the epidemiological model to calibrate on both reported infections $D_{\mathrm{reported}}$ and candidate unreported infections $D_{\mathrm{unreported}}$. We compute the newly reported infections and unreported infections as follows:

\begin{enumerate}
    \item New reported infections$ = \frac{\alpha P}{D_{p}}$: $\frac{P}{D_{p}}$ represents the number of new infections from presymptomatic infections every day in $O_{\mathrm{M}}$. Here, we assume $\alpha$ proportion of new infections every day will be that day's new reported infections.
    \item New unreported infections $ = \frac{(1-\alpha) P}{D_{p}}$: Then, the $1-\alpha$ proportion of new infections every day will be that day's new unreported infections.
\end{enumerate}

\subsection{SEIR+HD Model}

\begin{figure}[h]
    \centering
    \includegraphics[width=0.5\textwidth]{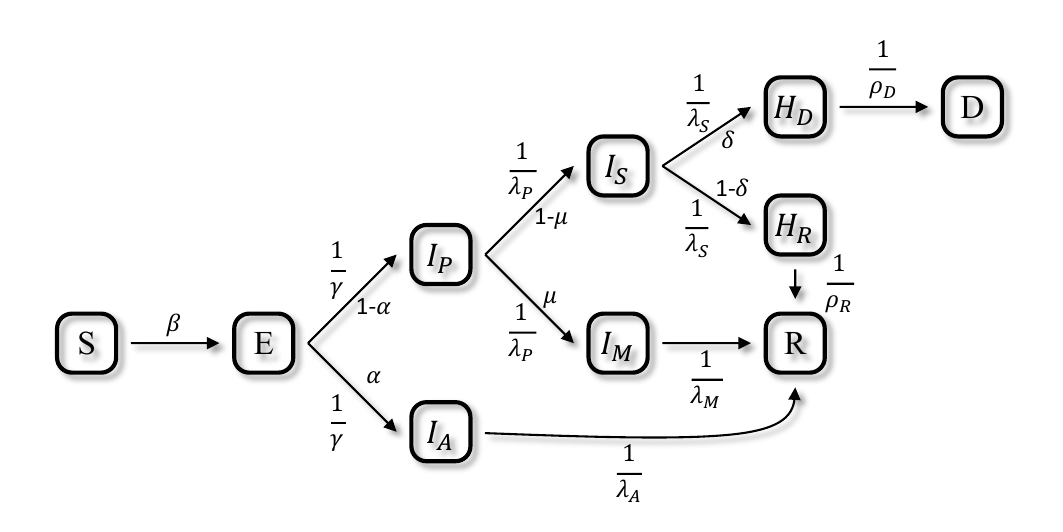}
    \caption{Compartmental diagram of SEIR+HD model~\cite{hao2020reconstruction}.}
    \label{fig:MordecaiDiagram}
\end{figure}

We also use the SEIR+HD model~\cite{kain2021chopping} as another epidemiological model $O_{\mathrm{M}}$ in our experiments. The compartmental diagram of SEIR+HD model is shown in Figure~\ref{fig:MordecaiDiagram}. The SEIR+HD model has of 21 different parameters. Note that only three parameters are calibrated, while the rest are fixed. 

Similarly to SAPHIRE model, we still expect the epidemiological model to calibrate on reported infections $D_{\mathrm{reported}}$ and candidate unreported infections $D_{\mathrm{unreported}}$. Hence we extend its calibration procedure to infer two more parameters: $\alpha$ and $\alpha_{1}$ (proportion of new symptomatic infections that are reported). We compute the newly reported infections and unreported infections as follows:

\begin{enumerate}
    \item New reported infections$ = \alpha_{1} \times (N_{I_{P}I_{S}} + N_{I_{P}I_{M}})$:\par
    $I_{\mathrm{new~sympt}} = N_{I_{P}I_{S}}+N_{I_{P}I_{M}}$ represents the number of new symptomatic infections every day in $O_{\mathrm{M}}$. Here, we assume $\alpha_{1}$ proportion of new symptomatic infections every day will be that day's new reported infections.
    \item New unreported infections $ = (1 - \alpha_{1}) \times (N_{I_{P}I_{S}} + N_{I_{P}I_{M}}) + N_{EI_{A}}$:\par
    Then, the $1-\alpha_{1}$ proportion of new symptomatic infections every day and new asymptomatic infections every day will be that day's new unreported infections.
\end{enumerate}

\subsection{Baseline Parametrization}
By calibrating $O_{\mathrm{M}}$ on $D_{\mathrm{reported}}$, we get the baseline paramterization $\Vector{p}$:
\begin{equation*}
    \Vector{p} = \Calibrate(O_{\mathrm{M}}, \{D_{\mathrm{reported}}, \mathrm{others}\})
    \label{eq:CalibrateP}
\end{equation*}

By running the epidemiological model with $\Vector{p}$, $O_{\mathrm{M}}$ will output the estimated reported infections $D_{\mathrm{reported}}(\Vector{p})$, estimated unreported infections $D_{\mathrm{unreported}}(\Vector{p})$, and estimated total infections $D(\Vector{p}) = D_{\mathrm{reported}}(\Vector{p}) + D_{\mathrm{unreported}}(\Vector{p})$. We can also calculate the reported rate $\alpha_{\mathrm{reported}}$ as follows:
\begin{equation*}
    \alpha_{\mathrm{reported}} = \frac{\sum D_{\mathrm{reported}}(\Vector{p})}{\sum D(\Vector{p})}
\end{equation*}
Here, we sum over the daily sequence $D_{\mathrm{reported}}(\Vector{p})$ and $D(\Vector{p})$ to calculate a scalar as the reported rate for MDL formulation.

\subsection{$\MDLFramework$ Parametrization}
Similarly by calibrating $O_{\mathrm{M}}$ on $D_{\mathrm{reported}}$ and $D_{\mathrm{unreported}}$, we get the candidate paramterization $\Vector{p'}$:
\begin{equation*}
    \Vector{p'} = \Calibrate(O_{\mathrm{M}}, \{D_{\mathrm{reported}}, D_{\mathrm{unreported}},\mathrm{others}\})
    \label{eq:CalibratePP}
\end{equation*}
By running the epidemiological model with $\Vector{p'}$, $O_{\mathrm{M}}$ will output the estimated reported infections $D_{\mathrm{reported}}(\Vector{p'})$, estimated unreported infections $D_{\mathrm{unreported}}(\Vector{p'})$, and estimated total infections $D(\Vector{p'}) = D_{\mathrm{reported}}(\Vector{p'}) + D_{\mathrm{unreported}}(\Vector{p'})$. Similarly, we can calculate the reported rate $\alpha_{\mathrm{reported}}'$ as follows:
\begin{equation*}
    \alpha_{\mathrm{reported}}' = \frac{\sum D_{\mathrm{reported}}(\Vector{p'})}{\sum D(\Vector{p'})}
\end{equation*}
With the calibration process, $\Vector{p}$, and $\Vector{p'}$ defined, we can next formalize the MDL cost.

\section{Methodology}
\subsection{Sender-receiver Framework}
%Here, we will use a two-part sender-receiver framework based on the Minimum Description Length (MDL) principle. In the framework, there exists four concepts: Sender \emph{S}, receiver \emph{R}, $\textsc{Model}$, and $\textsc{Data}$. The goal of the framework is to transmit the \textsc{Data} from the possession of sender \emph{S} to the receiver \emph{R} using a \textsc{Model}. We do this by identifying the \textsc{Model} describes the \textsc{Data} such that the total number of bits needed to represent both the \textsc{Model} and the \textsc{Data} is minimized. We term this as a cost function composed by two parts:

Here, we use the two-part sender-receiver framework based on the Minimum Description Length (MDL) principle. The goal of the framework is to transmit the \textsc{Data} in possession of the Sender \emph{S} to the receiver \emph{R} using a \textsc{Model}. We do this by identifying the \textsc{Model} that describes the \textsc{Data} such that the total number of bits needed to encode both the \textsc{Model} and the \textsc{Data} is minimized. The number of bits required to encode both the \textsc{Model} and the \textsc{Data} is given by the cost function $L$, which has two components: (i) model cost $L(\textsc{Model})$: The cost in bits of encoding the $\textsc{Model}$, and (ii) data cost $L(\textsc{Data}|\textsc{Model})$: The cost in bits of encoding $\textsc{Data}$ given the $\textsc{Model}$.

%We will introduce model space, $L(\textsc{Model})$, and $L(\textsc{Data}|\textsc{Model})$ in detail in the following sections:
\subsection{Model Space: Other Choice}
In this work, the $\textsc{Data}$ is $D_{\mathrm{reported}}$. One idea for defining the $\textsc{Model}$ space is to use  $\Vector{p}$. With such a $\textsc{Model}$, the receiver \emph{R} can easily compute first $D_{\mathrm{reported}}(\Vector{p})$ given $\Vector{p}$. Then the sender \emph{S} will only need to encode and send the difference between $D_{\mathrm{reported}}(\Vector{p})$ and $D_{\mathrm{reported}}$ so that the receiver can recover the $\textsc{Data}$ fully. However, this has the disadvantage that slightly different \Vector{p} could lead to vastly different $D_{\mathrm{reported}}(\Vector{p})$, and so the optimization problem will become hard to solve. To account for this, we propose $\textsc{Model}$ as $\textsc{Model} = (D, \Vector{p'}, \Vector{p})$ as described in the main article, which consists of three components.

\subsection{Model Cost}
With the model space $\textsc{Model} = (D, \Vector{p'}, \Vector{p})$, the sender \emph{S} will send the $\textsc{Model}$ to the receiver \emph{R} in three parts: (i) first send $\Vector{p}$, (ii) next send $\Vector{p'}$ given $\Vector{p}$, and then (iii) send $D$ given $\Vector{p'}$ and $\Vector{p}$. Therefore, the model cost $L(D,\Vector{p'},\Vector{p})$ will also have three components
\begin{equation*}
    L(D,\Vector{p'},\Vector{p}) = \Cost(\Vector{p}) + \Cost(\Vector{p'}|\Vector{p}) + \Cost(D|\Vector{p'},\Vector{p})
    \label{eq:ModelCost1}
\end{equation*}
Here, we will send the first component, $\Vector{p}$, directly, send the second component, $\Vector{p'}$ given $\Vector{p}$, via sending $\Vector{p'}-\Vector{p}$, and send the third component, $D$ given $\Vector{p'}$ and $\Vector{p}$, via sending $\alpha_{\mathrm{reported}}' \times D-D_{\mathrm{reported}}(\Vector{p})$. We further write the model cost as below:
\begin{equation*}
    \begin{aligned}
        L(D,\Vector{p'},\Vector{p}) &= \Cost(\Vector{p}) + \Cost(\Vector{p'}-\Vector{p}|\Vector{p}) \\ &+ \Cost(\alpha_{\mathrm{reported}}' \times D-D_{\mathrm{reported}}(\Vector{p})|\Vector{p'},\Vector{p})
    \end{aligned}
    \label{eq:ModelCost2}
\end{equation*}

\subsection{Data Cost}
Give the $\textsc{Model} = (D, \Vector{p'}, \Vector{p})$ and model cost above, next we will send the $\textsc{Data}$ in terms of the $\textsc{Model}$. Here, the $\textsc{Data}$ is $D_{\mathrm{reported}}$, and the data cost will have only one component:
\begin{equation*}
    L(D_{\mathrm{reported}}|D,\Vector{p'},\Vector{p}) = \Cost(D_{\mathrm{reported}}|D,\Vector{p'},\Vector{p})
    \label{eq:DataCost1}
\end{equation*}
Here, we will send it via $\frac{D-D_{\mathrm{reported}}}{1-\alpha_{\mathrm{reported}}'}-D(\Vector{p'})$, and we further write the data cost as below:

\begin{equation*}
\scalebox{0.8}{
    $L(D_{\mathrm{reported}}|D,\Vector{p'},\Vector{p}) = \Cost(\frac{D-D_{\mathrm{reported}}}{1-\alpha_{\mathrm{reported}}'}-D(\Vector{p'})|D,\Vector{p'},\Vector{p})$}
    \label{eq:DataCost2}
\end{equation*}

\subsection{Total Cost}
The total cost is the sum of model cost $L(D,\Vector{p'},\Vector{p})$ and data cost $L(D_{\mathrm{reported}}|D,\Vector{p'},\Vector{p})$:
\begin{equation*}
    \scalebox{0.8}{
    $\begin{aligned}
        L(D_{\mathrm{reported}},D,\Vector{p'},\Vector{p}) &= L(D,\Vector{p'},\Vector{p}) + L(D_{\mathrm{reported}}|D,\Vector{p'},\Vector{p}) \\ 
        &= \Cost(\Vector{p}) + \Cost(\Vector{p'}|\Vector{p}) \\ &+ \Cost(D|\Vector{p'},\Vector{p}) + \Cost(D_{\mathrm{reported}}|D,\Vector{p'},\Vector{p}) \\
        &= \Cost(\Vector{p}) + \Cost(\Vector{p'}-\Vector{p}|\Vector{p}) \\ &+ \Cost(\alpha_{\mathrm{reported}}' \times D-D_{\mathrm{reported}}(\Vector{p})|\Vector{p'},\Vector{p}) \\ & + \Cost(\frac{D-D_{\mathrm{reported}}}{1-\alpha_{\mathrm{reported}}'}-D(\Vector{p'})|D,\Vector{p'},\Vector{p}) 
    \end{aligned}$}
    \label{eq:MDL2}
\end{equation*}

%Here, the data is the $D_{\mathrm{reported}}$, which is in process of the sender. The model is the the total infection line-lists $D$. The goal of the framework is to transmit the $D_{\mathrm{reported}}$ ($\textsc{Data}$) from the sender to receiver, and we will send the $D_{\mathrm{reported}}$ via sending the total infection line-lists $D$ ($\textsc{Model}$). 

\subsection{Cost Derivation}

Next, we derive the cost for each component and give our encoding method explicitly:

\begin{enumerate}
    \item $\Cost(\Vector{p})$: We represent $\Vector{p}$ as a vector of real numbers (we describe our encoding later below). 
    \item  $\Cost(\Vector{p'}-\Vector{p}|\Vector{p})$: We will encode the difference of two vectors as a vector of real numbers.
    \item $\Cost(\alpha_{\mathrm{reported}}' \times D-D_{\mathrm{reported}}(\Vector{p})|\Vector{p'},\Vector{p})$:  Here, we encode the difference between the two time sequences: $\alpha_{\mathrm{reported}}' \times D$ given $D_{\mathrm{reported}}(\Vector{p})$.
    \item  $\Cost(\frac{D-D_{\mathrm{reported}}}{1-\alpha_{\mathrm{reported}}'}-D(\Vector{p'})|D,\Vector{p'},\Vector{p})$: Again, we encode it as a difference between the two time sequences: $\frac{D_{\mathrm{unreported}}}{1-\alpha_{\mathrm{reported}}'}$ given $D(\Vector{p'})$.
\end{enumerate}
Next, we describe the encoding cost of real numbers, vectors, and the difference between two time sequences.\par
\subsubsection{Encoding Integers}
%Before introducing the encoding method for real numbers, we need to first introduce the encoding method for integers.\par
To encode a positive integer $n$, we encode both the binary representation of integer $n$ as well as the length of the representation $\log_{2}n$. Following~\cite{lee2001introduction}, we use the cost in bits of encoding a single integer $n$ is as follows:
\begin{equation*}
    \Cost(n) = \log_{2}c_{0} + \log^{*}(n).
\end{equation*}
where $c_{0} \approx 2.865$ and $\log^{*}(n)= \log_{2}n + \log_{2}\log_{2}n + \cdots $ as described in~\cite{lee2001introduction}. There are infinite terms in $\log^{*}(n)$ function since after we encoded a number, we always need to encode its length as another number, which could be repeated for infinite times. Additionally, if we want to transmit an integer that can be either positive or negative, we can add another sign bit and therefore the cost in bits for integers will be
\begin{equation*}
    \Cost(n) = \Cost(|n|) + 1.
\end{equation*}
\subsubsection{Encoding Real Numbers}
Note that most real numbers (e.g. $\pi$ or $e$) need infinite number of bits to encode. Hence, we introduce a precision threshold $\delta$. With threshold $\delta$, we approximate a real number $x$ with $x_{\delta}$ which satisfies $|x-x_{\delta}|<\delta$, and we encode $x_{\delta}$ instead. To encode $x_{\delta}$, we encode both the integer part $\lfloor x \rfloor$ as well as the fractional part $x_{\delta}-\lfloor x \rfloor$. Hence the cost in bits of encoding a real number $x$ is as follows:

\begin{equation*}
    \Cost(x) = \Cost(\lfloor x \rfloor) + \log_{2}\frac{1}{\delta}
\end{equation*}
where $\lfloor x \rfloor$ is the floor of $x$ and therefore is a integer, whose encoding cost is $\Cost(\lfloor x \rfloor) = \log_{2}c_{0} + \log^{*}(\lfloor x \rfloor)$. Additionally, if we want to transmit a real number that can be either positive or negative, we can add another sign bit and therefore the cost in bits for real numbers will be
\begin{equation*}
    \Cost(x) = \Cost(|x|) + 1
\end{equation*}
%Specifically, if we encode a real number in the range of $-1$ to $1$, then the cost will be a fixed number $\Cost(0) + \log_{2} \frac{1}{\delta}$, which is a constant value.\par
\subsubsection{Encoding Vectors}
To encode a vector $\Vector{p}=[\Vector{p}[1],\Vector{p}[2],\cdots,\Vector{p}[n]]$, we encode every components one by one as real numbers. Hence the cost in bits of encoding a vector $\Vector{p}$ is as follows:
\begin{equation*}
    \Cost(\Vector{p}) = \Cost(\Vector{p}[1]) + \Cost(\Vector{p}[2])+\cdots+ \Cost(\Vector{p}[n])
\end{equation*}
\subsubsection{Encoding The Difference between Two Time Sequences}
To encode the difference $A-B = [A_{t_{1}}-B_{t_{1}}, A_{t_{2}}-B_{t_{2}}, \cdots, A_{t_{n}}-B_{t_{n}}]$ between two time sequence $A = [A_{t_{1}}, A_{t_{2}}, \cdots, A_{t_{n}}]$ and $B = [B_{t_{1}}, B_{t_{2}}, \cdots, B_{t_{n}}]$, we encode every components one by one as real numbers. Hence the cost in bits of encoding the difference is as follows:
\begin{equation*}
\begin{aligned}
    \Cost(A-B) &= \Cost(A_{t_{1}}-B_{t_{1}}) + \Cost(A_{t_{2}}-B_{t_{2}})\\ &+ \cdots + \Cost(A_{t_{n}}-B_{t_{n}})
\end{aligned}
\end{equation*}

\subsection{Problem Statement}
%With MDL cost function $L(D_{\mathrm{reported}},D,\Vector{p'},\Vector{p})$ formulated above, we can state the problem of searching for the best total infections $D^{*}$ as follows:\par
Now we have derived every cost involved in our problem, and we can finally state our problem as one of estimating the total infections $D$ as follows: Given the time sequence $D_{\mathrm{reported}}$ and epidemiological model $O_{\mathrm{M}}$, find $D^{*}$ that minimizes the MDL total cost:
\begin{equation*}
    D^{*} = \arg \min_{D} L(D_{\mathrm{reported}},D,\Vector{p'},\Vector{p})
\end{equation*}
We will give the algorithm to find such $D^{*}$ as follows:

\begin{figure}[ht]
\centering
\includegraphics[width=0.5\textwidth]{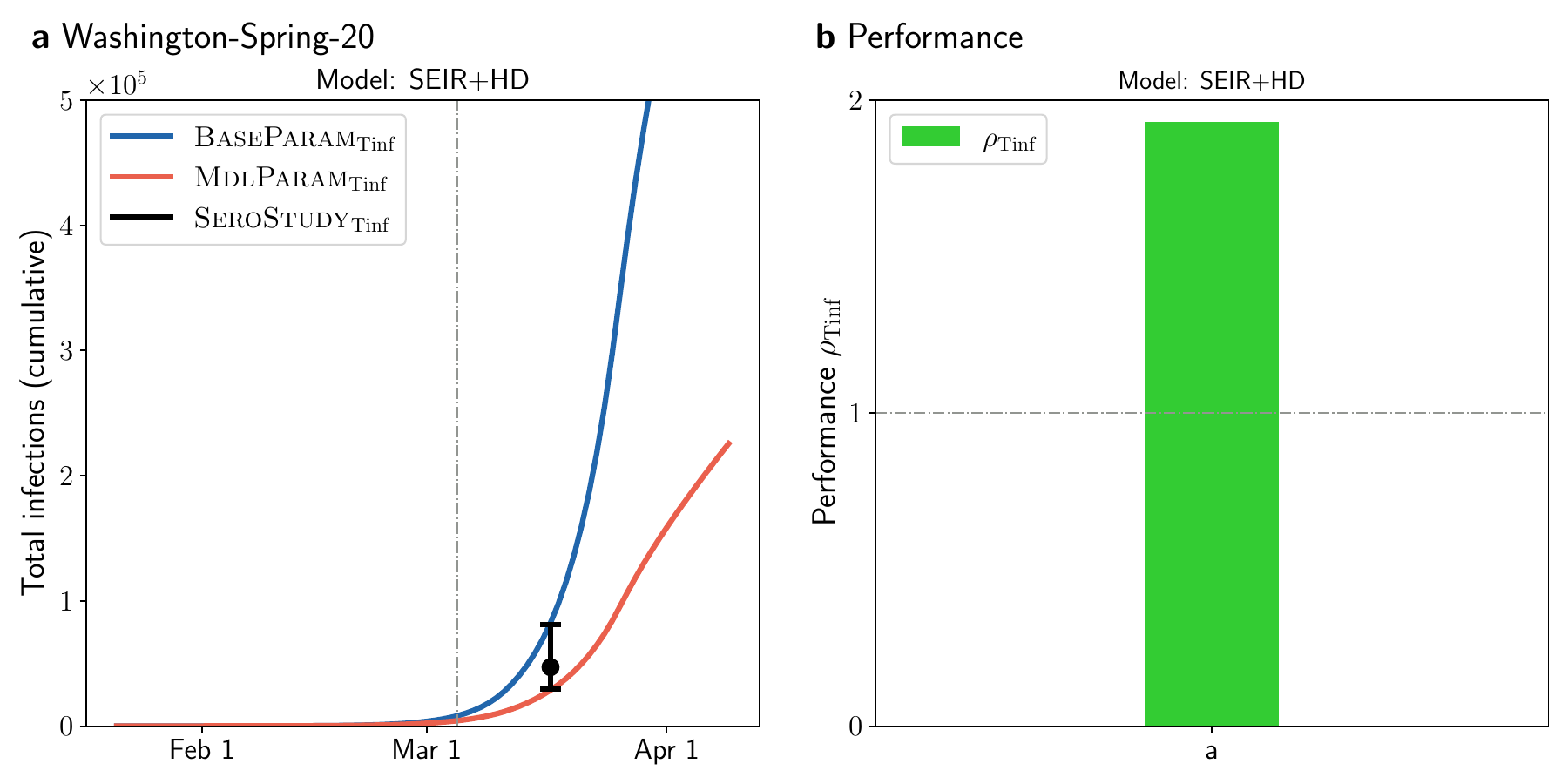}
\caption{$\MDLResult$ (red) gives a closer estimation of total infections to serological studies (black) than $\MordecaiResult$ (blue). Note that the serological studies are not informed for both approaches. \textbf{a} The red and blue curves represent $\MDLFramework$'s estimation of total infections, $\MDLResult_{\mathrm{Tinf}}$, and baseline calibration procedure's estimation of total infections, $\MordecaiResult_{\mathrm{Tinf}}$, respectively. The black point estimates and confidence intervals represent the total infections estimated by serological studies~\cite{CDCTracker, havers2020seroprevalence}, or $\textsc{SeroStudy}_{\mathrm{Tinf}}$. \textbf{b} The performance metric, $\rho_{\mathrm{Tinf}}$, comparing $\MDLResult_{\mathrm{Tinf}}$ against $\MordecaiResult_{\mathrm{Tinf}}$ is shown for \textbf{a} for the SEIR+HD model. Here, the values of $\rho_{\mathrm{Tinf}}$ is 1.93.}
\label{Serological}
\end{figure}

\subsection{Algorithms}
Before presenting our algorithm to find $D^{*}$, we will first address the problem of searching $D^{*}$ directly. Note that $D^{*}$ is a time sequence of total infections instead of a scalar, naively searching $D^{*}$ directly in large search space is intractable. Hence, we propose an alternate method: First, we can find quickly a good reported rate $\alpha_{\mathrm{reported}}^{*}$ since we can constrain $D = \frac{D_{\mathrm{reported}}}{\alpha_{\mathrm{reported}}}$ to reduce the search space. Then we can search for the optimal $D^{*}$ with $\alpha_{\mathrm{reported}}^{*}$ from step 1 as constraints. Here, we write down our two-step search algorithm to find the $D^{*}$ as follows:
\begin{enumerate}
    \item Step 1: We do a linear search to find a good reported rate $\alpha_{\mathrm{reported}}^{*}$, which serves as an initialization in the second step.
    \item Step 2: Given the $\alpha_{\mathrm{reported}}^{*}$ found in step 1, we use the Nelder-Mead~\cite{gao2012implementing} optimization to find the $D^{*}$ that minimizes $L(D_{\mathrm{reported}},D,\Vector{p'},\Vector{p})$ with $\alpha_{\mathrm{reported}}^{*}$ constraints.
\end{enumerate}
\subsubsection{Step 1: Find the $\mathbf{\alpha_{\mathrm{reported}}^{*}}$}
In step 1, we search on $\alpha_{\mathrm{reported}}$ to find the $\alpha_{\mathrm{reported}}^{*}$ as follows:
\begin{equation*}
    \alpha_{\mathrm{reported}}^{*} = \arg \min_{\alpha_{\mathrm{reported}}} L(D_{\mathrm{reported}},\frac{D_{\mathrm{reported}}}{\alpha_{\mathrm{reported}}}, \Vector{p'},\Vector{p})
\end{equation*}
To be more specific, in the first step of our algorithm, we do a linear search on different $\alpha_{\mathrm{reported}} = [0.01, 0.02, 0.03, \cdots,0.99]$ and calibrate the $O_{\mathrm{M}}$ on $D=\frac{D_{\mathrm{reported}}}{\alpha_{\mathrm{reported}}}$, which means

\begin{equation*}
\scalebox{0.8}{
    $\Vector{p'} = \Calibrate(O_{\mathrm{M}}, \{D_{\mathrm{reported}},\frac{D_{\mathrm{reported}}}{\alpha_{\mathrm{reported}}}-D_{\mathrm{reported}}, \mathrm{others}\})
    $}
\end{equation*}

Then we pick the $\alpha_{\mathrm{reported}}^{*}$ that corresponds to the lowest total cost $L(D_{\mathrm{reported}}, D, \Vector{p'}, \Vector{p})$ as the $\mathbf{\alpha_{\mathrm{reported}}^{*}}$.
%Here, the reason why we use $D(\Vector{p'})$ instead of $\frac{D_{\mathrm{reported}}}{\alpha_{\mathrm{reported}}}$ as $D$ in the MDL cost in step 1 algorithm is that the $\frac{D_{\mathrm{reported}}}{\alpha_{\mathrm{reported}}}$ magnifies the noise of reported infections by $\frac{1}{\alpha_{\mathrm{reported}}}$ times. Hence smaller $\alpha_{\mathrm{reported}}$ tends to magnify the noise larger and leads to a bias of higher MDL cost. To cancel this bias introduced by the noise, we use the more smooth $D(\Vector{p'})$ as $D$ in step 1. \par
%Specifically, we will use $D(\Vector{p'})$ instead of $\frac{D_{\mathrm{reported}}}{\alpha_{\mathrm{reported}}}$ as $D$ when measuring the total cost $L(D_{\mathrm{reported}},D, \Vector{p'},\Vector{p})$. The reason behind it is that $\frac{D_{\mathrm{reported}}}{\alpha_{\mathrm{reported}}}$ magnifies the noise of reported infections by $\frac{1}{\alpha_{\mathrm{reported}}}$ times. Hence smaller $\alpha_{\mathrm{reported}}$ tends to magnify the noise larger and leads to a bias towards higher total cost. To cancel this bias introduced by the noise and get stable and robust $\alpha_{\mathrm{reported}}$, we use the more smooth $D(\Vector{p'})$ as $D$ in step 1. \par

\subsubsection{Step 2: Find the $D^{*}$ given $\alpha_{\mathrm{reported}}^{*}$}
With $\alpha_{\mathrm{reported}}^{*}$ inferred in step 1, we will next find the $D^{*}$ that minimizes the total cost. 
\begin{equation*}
    D^{*} = \arg \min_{D} L(D_{\mathrm{reported}}, D, \Vector{p'}, \Vector{p})
\end{equation*}
Since we have already found $\alpha_{\mathrm{reported}}^{*}$ in step 1, we will only search the $D^{*}$ that satisfies
\begin{equation*}
    \sum D^{*} = \frac{\sum{D_{\mathrm{reported}}}}{\alpha_{\mathrm{reported}}^{*}}
\end{equation*}
To search for the optimal $D^*$, we leverage the popular Nelder-Mead search algorithm~\cite{gao2012implementing}. % $\frac{D_{\mathrm{reported}}}{\alpha_{\mathrm{reported}}^{*}}$.

\section{Metrics}

To evaluate the performance of $\MDLFramework$, as mentioned in the main article, we use the Root Mean
Squared Error (RMSE) following previous works~\cite{rodriguez2023einns,rodriguez2021deepcovid,cramer2022evaluation}. Specifically, the metrics are calculated using the following equataions following~\cite{roy2021deep}.

\begin{equation*}
    RMSE = \sqrt{\sum_{t}(\hat{y}_{t}-y_{t})^2}
\end{equation*}

where $\hat{y}_t$ is the estimated number by either $\BaseInfer$ or $\MDLFramework$, $y_t$ is the ground-truth value.

\section{Experimental Setup}
Here we describe our experimental setup in more detail and present results on additional testbeds. We also list the notations used in the experiments section in Table~\ref{table:S1}.

\subsection{Total Infections}
The Results section in the main paper refers to $\MordecaiResult_{\mathrm{Tinf}}$, which represents the cumulative total infections derived from the baseline calibration procedure. It is computed as follows:
\begin{equation*}
    \MordecaiResult_{\mathrm{Tinf}} = \sum D(\Vector{p})
\end{equation*}

Similarly, $\MDLResult_{\mathrm{Tinf}}$, which represents the cumulative total infections derived from $\MDLFramework$, is computed as follows:
\begin{equation*}
    \MDLResult_{\mathrm{Tinf}} = \sum D(\Vector{p'})
\end{equation*}

\begin{figure*}[ht]
\centering
\includegraphics[width=\textwidth]{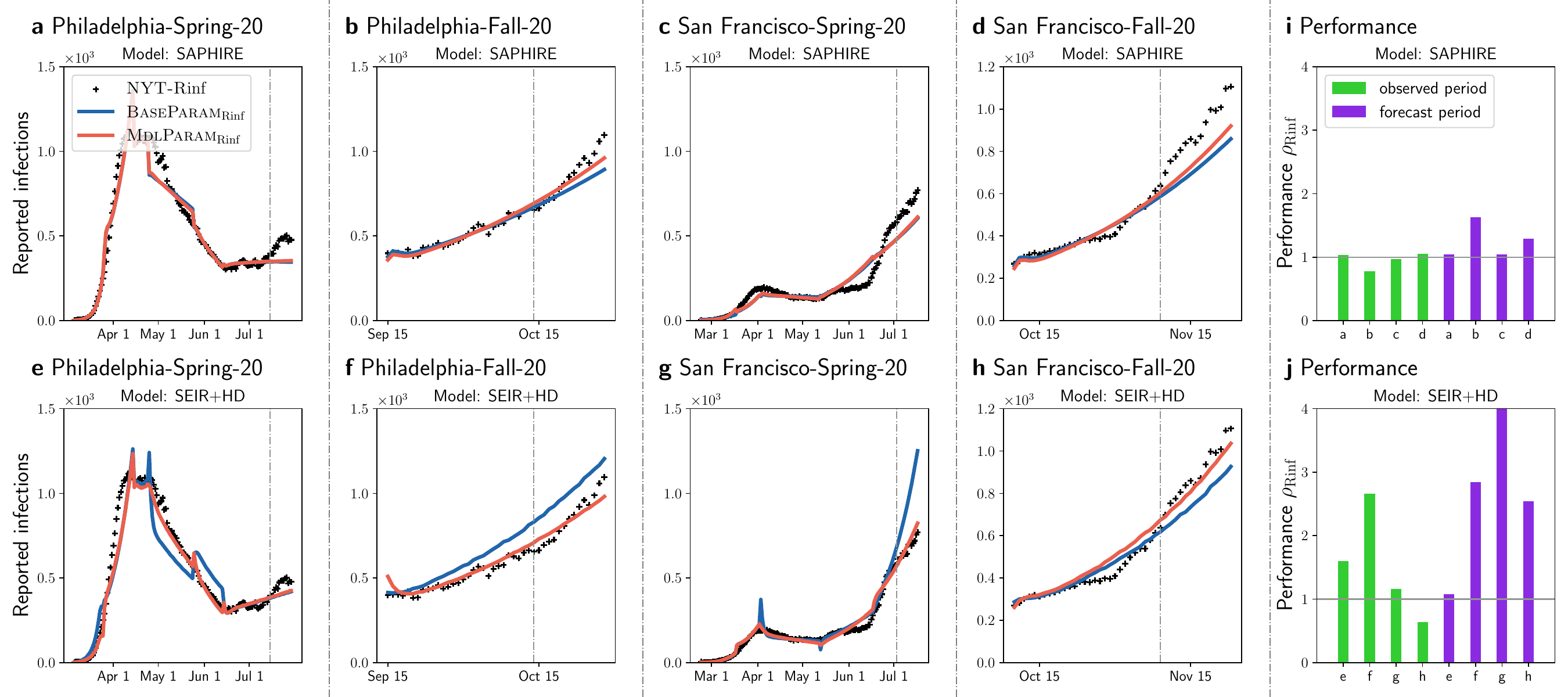}
\caption{$\MDLResult$ (red) gives a closer estimation of reported infections (black) than $\MordecaiResult$ (blue) on various geographical regions and time periods. We use the reported infections in the observed period as inputs and try to forecast the future reported infections (forecast period). \textbf{a}-\textbf{h} The vertical grey dash line divides the observed period and forecast period. The red and blue curves represent $\MDLFramework$'s estimation of reported infections, $\MDLResult_{\mathrm{Rinf}}$, and baseline calibration procedure's estimation of reported infections, $\MordecaiResult_{\mathrm{Rinf}}$, respectively. The black plus symbols represent the infections reported by the New York Times ($\textsc{NYT-R}\mathrm{inf}$). \textbf{a}-\textbf{d} is for SAPHIRE model and \textbf{e}-\textbf{h} is for SEIR+HD model. \textbf{i}-\textbf{j} The performance metric, $\rho_{\mathrm{Rinf}}$, comparing $\MDLResult_{\mathrm{Rinf}}$ against $\MordecaiResult_{\mathrm{Rinf}}$ is shown for \textbf{a}-\textbf{h} for the regions for both the SAPHIRE model in \textbf{i}, and the SEIR+HD model in \textbf{j}.}
\label{Reported}
\end{figure*}

In Figure~\ref{Serological}, we show additional results comparing the performance of $\MDLFramework$ and baseline calibration procedure in estimating total infections. Here, $\MDLFramework$ (red) gives a closer estimation of total infections to serological studies (black) than baseline calibration procedure (blue).

\subsection{Reported Infections}

In Figure~\ref{Reported}, we present additional results comparing the performance of $\MDLFramework$ and baseline calibration procedure in forecasting future infections (forecast period). Here, $\MDLFramework$ (red) gives a closer estimation of reported infections (black) than baseline calibration procedure (blue) on various geographical regions and time periods.

\subsection{Symptomatic Rate}
The baseline calibration procedure and $\MDLFramework$ also estimate the symptomatic rate $\MordecaiResult_{\mathrm{Symp}}$ and $\MDLResult_{\mathrm{Symp}}$ respectively. We compare these against the Facebook symptomatic surveillance data $\textsc{Rate}_{\mathrm{Symp}}$. 

We calculate $\MordecaiResult_{\mathrm{Symp}}$ from $\Vector{p}$ as follows:
\begin{equation*}
    \MordecaiResult_{\mathrm{Symp}} = \frac{I_{S}(\Vector{p})+I_{M}(\Vector{p})}{N}
\end{equation*}
where $I_{S}(\Vector{p})$ is the number of infections in severe symptomatic state, $I_{M}(\Vector{p})$ represents the same in mild symptomatic state, and $N$ is the total population in this area.\par

\begin{figure*}[!htbp]
\centering
\includegraphics[width=\textwidth]{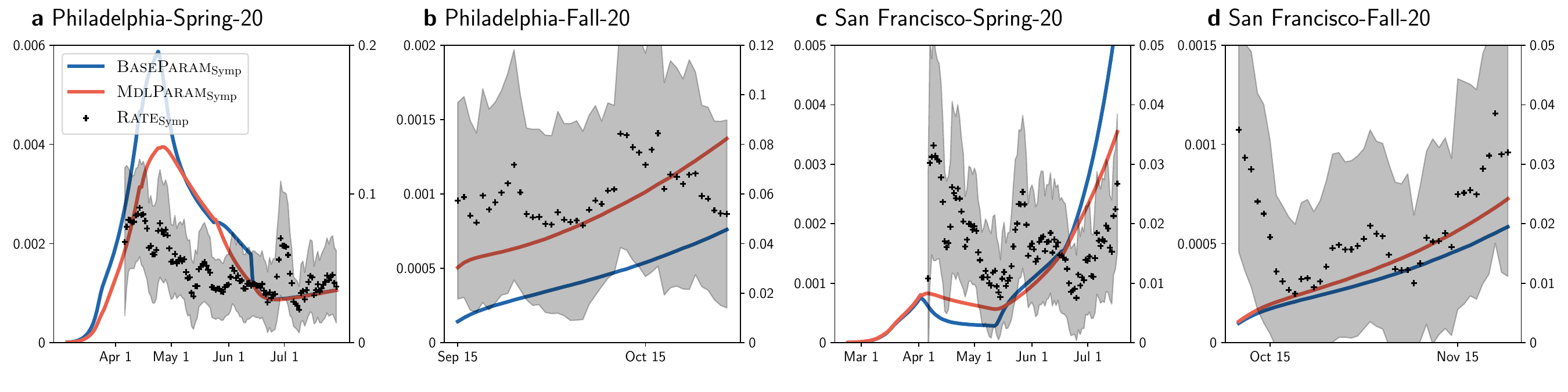}
\centering
\caption{$\MDLResult$ (red) gives a closer estimation of symptomatic rate (black) than $\MordecaiResult$ (blue). The red and blue curves represent $\MDLFramework$'s estimation of symptomatic rate, $\MDLResult_{\mathrm{Symp}}$, and baseline calibration procedure's estimation of symptomatic rate, $\MordecaiResult_{\mathrm{Symp}}$, respectively. The black points and the shaded regions are the point estimate with standard error for $\textsc{Rate}_{\mathrm{Symp}}$ (the COVID-related symptomatic rates derived from the symptomatic surveillance dataset~\cite{delphisurvey,salomon2021us}).}
\label{Symptomatic}
\end{figure*}

Similarly $\MDLResult_{\mathrm{Symp}}$ is computed as follows:
\begin{equation*}
    \MDLResult_{\mathrm{Symp}} = \frac{I_{S}(\Vector{p'})+I_{M}(\Vector{p'})}{N}
\end{equation*}

In Figure~\ref{Symptomatic}, we present additional results comparing $\MDLFramework$ and baseline calibration procedure in estimating trends of symptomatic rate. Here, $\MDLFramework$ (red) gives a closer estimation of symptomatic rate (black) than baseline calibration procedure (blue).

\subsection{Cumulative Reported Rate}
We also calculate the a dynamic reported rate from both baseline calibration procedure and $\MDLFramework$. Note that this cumulative reported rate is different from $\alpha_{\mathrm{reported}}$ and $\alpha_{\mathrm{reported}}'$, which are two scalars used in MDL formulation. We calculate $\MordecaiResult_{\mathrm{Rate}}$ from $\MordecaiResult$ $\Vector{p}$ as follows:
\begin{equation*}
    \MordecaiResult_{\mathrm{Rate}} = \frac{\sum \textsc{NYT-R}\mathrm{inf}}{\sum D(\Vector{p})}
\end{equation*}
%which are the cumulative numbers from the start of the epidemic.\par
Similarly we calculate $\MDLResult_{\mathrm{Rate}}$ from $\MDLResult$ $\Vector{p'}$ as follows:
\begin{equation*}
    \MDLResult_{\mathrm{Rate}} = \frac{\sum  \textsc{NYT-R}\mathrm{inf}}{\sum D(\Vector{p'})}
\end{equation*}

\par

\subsection{Non-pharmaceutical Interventions Simulation}
We also use the baseline calibration procedure and $\MDLFramework$ to perform non-pharmaceutical interventions simulation on SEIR+HD model. Here, both the baseline calibration procedure and $\MDLFramework$ are estimated on the observed period, then on the future period, we will consider the following five scenarios of isolation:
\begin{enumerate}
    \item Isolate reported infections: We isolate the $\alpha_{1}$ fraction of severe symptomatic infections $I_{S}$ and mild symptomatic infections $I_{M}$.
    \item Isolate both reported infections and symptomatic infections: Note that some reported infections are included in the symptomatic infections. Here, we isolate all severe symptomatic infections $I_{S}$ and mild symptomatic infections $I_{M}$. 
    \item Isolate 25\% presymptomatic and asymptomatic infections: We isolate 25\% of presymptomatic infections $I_{P}$, asymptomatic infections $I_{A}$, and all severe symptomatic infections $I_{S}$ and mild symptomatic infections $I_{M}$.
    \item Isolate 50\% presymptomatic and asymptomatic infections: We isolate 50\% of presymptomatic infections $I_{P}$, asymptomatic infections $I_{A}$, and all severe symptomatic infections $I_{S}$ and mild symptomatic infections $I_{M}$.
    \item Isolate 75\% presymptomatic and asymptomatic infections: We isolate 75\% of presymptomatic infections $I_{P}$, asymptomatic infections $I_{A}$, and all severe symptomatic infections $I_{S}$ and mild symptomatic infections $I_{M}$.
\end{enumerate}
The infectiousness of the noes in isolated is reduces by 50\%.
%where the isolation means to reduce the infectiousness of these infections to 50\% of the original value. \par 

\section{Sensitive Analysis}
\begin{figure*}[!p]
\centering
\vspace{-25em}
\includegraphics[width=\textwidth]{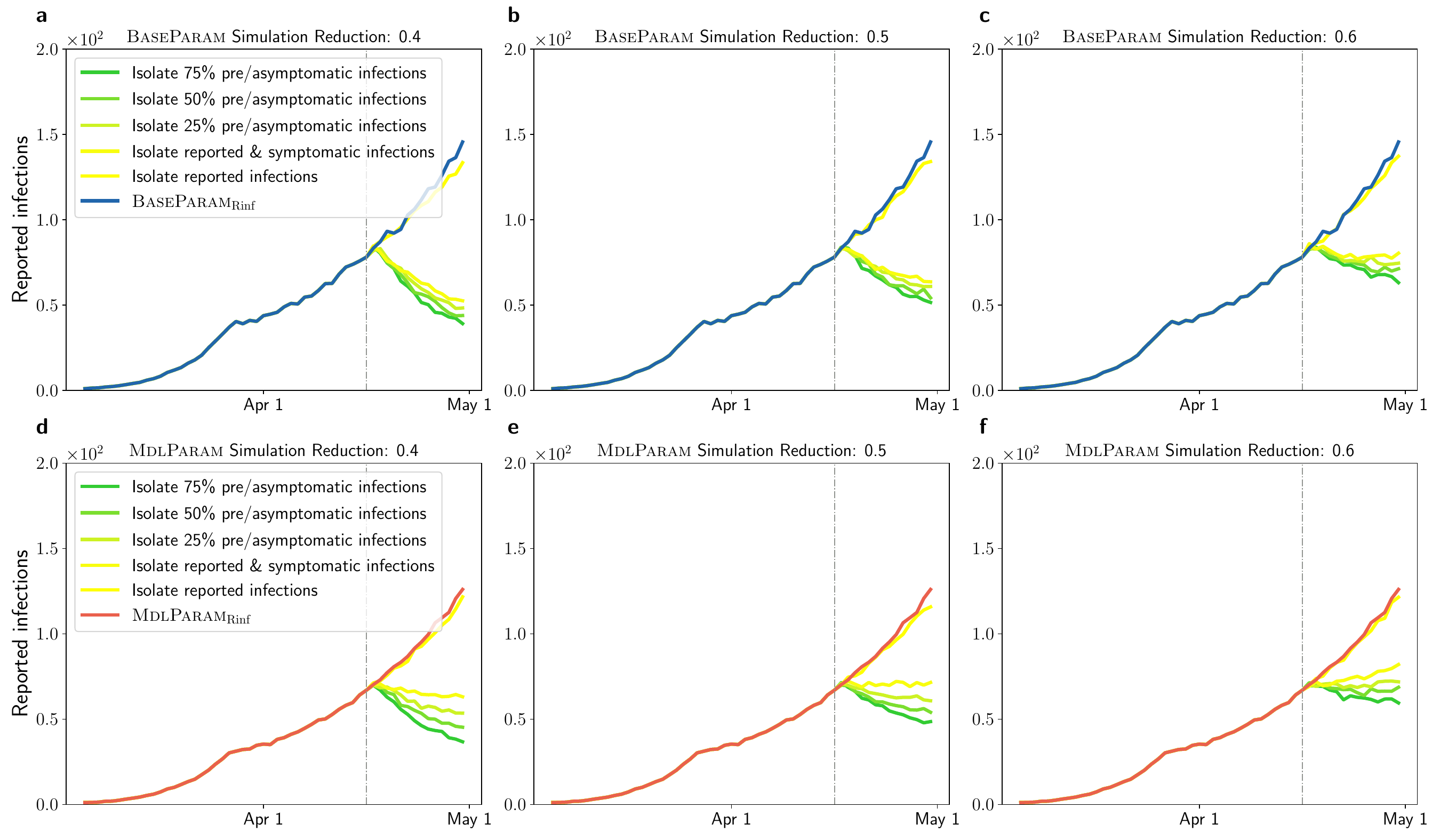}
\caption{Our non-pharmaceutical interventions simulation results are robust. \textbf{a}-\textbf{c}, the vertical grey dash line divides the observed period and future period. The blue curve represents the $\MordecaiResult$'s estimation of reported infections. The other five curves represent the simulated reported infections for 5 scenarios: (i) Isolate the reported infections, (ii) symptomatic infections, symptomatic infections and (iii) 25\%, (iv) 50\%, (v) 75\% asymptomatic and presymptomatic infections, where we reduce the infectiousness of these isolated infections to 40\% in \textbf{a}, 50\% in \textbf{b}, and 60\% in \textbf{c} in future period. \textbf{d}-\textbf{f}, the vertical grey dash line divides the observed period and future period. The red curve represents the $\MDLResult$'s estimation of reported infections. The other five curves represent the simulated reported infections for the same 5 scenarios as in \textbf{a} to \textbf{c}. The results are for Minneapolis-Spring-20 that we shown in main article Figure 5.}
\vspace{-2em}
\label{Hennepin}
\end{figure*}

We also perform sensitivity experiments to inspect the robustness of our non-pharmaceutical interventions simulations for Minneapolis-Spring-20 in Figure~\ref{Hennepin}. Here, we reduce the infectiousness of the isolated infections to 3 different values: 0.4, 0.5, and 0.6, and repeat simulations in each of the scenarios.

Our results consistently show that only isolating reported or symptomatic infections is not be enough to reduce the future reported infections. However, isolating both symptomatic infections and some fraction of asymptomatic and presymptomataic infections leads to reduction in reported infections in most settings.
\clearpage

\clearpage

\begin{table}[h!]
\centering
\caption{List of notations}
\label{table:S1}
\begin{tabular}{ll}
\toprule
Symbol & Description \\
\midrule
$\MDLFramework$ & Our Minimum Description Length (MDL) framework to estimate total infections \\
$O_{\mathrm{M}}$ & Epidemiological models used in $\MDLFramework$ \\
$\MordecaiResult$ & Baseline parameterization obtained via baseline calibration procedure\\
$\MDLResult$ & Optimal parametrization identified by $\MDLFramework$\\
$\textsc{SeroStudy}_{\mathrm{Tinf}}$ & Total infections estimated by serological studies \\
$\MordecaiResult_{\mathrm{Tinf}}$ & Total infections estimated by baseline calibration procedure \\
$\MDLResult_{\mathrm{Tinf}}$ & Total infections estimated by $\MDLFramework$ \\
$\rho_{\mathrm{Tinf}}$ & The performance metric comparing $\MDLFramework$ against \\
&baseline calibration procedure in estimating total infections \\
$\textsc{NYT-R}\mathrm{inf}$ & New York Times reported infections \\
$\MordecaiResult_{\mathrm{Rinf}}$ & Reported infections estimated by baseline calibration procedure \\
$\MDLResult_{\mathrm{Rinf}}$ & Reported infections estimated by $\MDLFramework$ \\
$\rho_{\mathrm{Rinf}}$ & The performance metric comparing $\MDLFramework$ against \\
& baseline calibration procedure in estimating reported infections \\
$\textsc{Rate}_{\mathrm{Symp}}$ & COVID-related symptomatic rate from symptomatic surveillance data \\
$\MordecaiResult_{\mathrm{Symp}}$ & Symptomatic rate estimated by baseline calibration procedure \\
$\MDLResult_{\mathrm{Symp}}$ & Symptomatic rate estimated by $\MDLFramework$ \\
$\textsc{SeroStudy}_{\mathrm{Rate}}$ & Cumulative reported rate estimated by serological studies \\
$\MordecaiResult_{\mathrm{Rate}}$ & Cumulative reported rate estimated by baseline calibration procedure \\
$\MDLResult_{\mathrm{Rate}}$ & Cumulative reported rate estimated by $\MDLFramework$ \\
\bottomrule
\end{tabular}
\end{table}

\end{document}